# Silicon photonic MEMS switches based on split waveguide crossings

Yinpeng Hu[1], Yi Sun[1], Ye Lu[1], Huan Li[1,2*], Liu Liu[1,2,3], Yaocheng Shi[1,2,3], and Daoxin Dai[1,2,3*]

[1]State Key Laboratory of Extreme Photonics and Instrumentation, Center for Optical & Electromagnetic Research, College of Optical Science and Engineering, International Research Center for Advanced Photonics, Zhejiang University, Zijingang Campus, Hangzhou 310058, China

[2]Jiaxing Key Laboratory of Photonic Sensing & Intelligent Imaging, Intelligent Optics & Photonics Research Center, Jiaxing Research Institute, Zhejiang University, Jiaxing 314000, China

[3]Ningbo Research Institute, Zhejiang University, Ningbo 315100, China

*e-mail: lihuan20@zju.edu.cn, dxdai@zju.edu.cn

## Abstract

The continuous push for high-performance photonic switches is one of the most crucial premises for the sustainable scaling of programmable and reconfigurable photonic circuits for a wide spectrum of applications. Conventional optical switches rely on the perturbative mechanisms of mode coupling or mode interference, resulting in inherent bottlenecks in their switching performance concerning size, power consumption and bandwidth. Here we propose and realize a silicon photonic 2×2 elementary switch based on a split waveguide crossing (SWX) consisting of two halves. The propagation direction of the incident light is manipulated to implement the OFF/ON states by splitting/combining the two halves of the SWX, showing excellent performance with low excess loss and low crosstalk over an ultrawide bandwidth. Both elementary switch and a 64×64 switch array based on Benes topology are fabricated and characterized, demonstrating great potential for practical scenarios such as photonic interconnect/routing, Lidar and spectroscopy, photonic computing, as well as microwave photonics.

**Introduction**

The rapid development of Artificial Intelligence (AI) and Internet of Things (IoT) are driving explosive demands for ultrahigh-capacity data transmission and signal processing[1]. Accordingly, it is required to develop sophisticated large-scale photonic integrated circuits (PICs) with high flexibility to be programmable and reconfigurable[2-4], where photonic switches of every scale (from 1×2 to M×N) are playing a key role to improve the utilization of the hardware resources as well as to reduce latency and energy consumption[5, 6]. Photonic switches have been demonstrated for various PICs in photonic interconnect/routing[7], Lidar[8] and spectroscopy[9], photonic computing[10], as well as microwave photonics[11, 12].

For example, large-scale photonic switches are indispensable for implementing agile, flexible, and scalable optical packet/burst switching (OPS/OBS) in datacenters. In this case, the scale of the photonic switches is preferably to be as large as 64×64[13]. For AI deep neural networks (DNN), photonic interconnect based on large-scale photonic switches enables massive data exchange among enormous amounts (tens of thousands or even more) of memories and processors, which avoids electronic interconnect with optical-electrical-optical (O-E-O) conversion and provides much broader bandwidths for data exchange[14]. The implementation of high-resolution focal-plane-switch-array (FPSA) lidars[15] also requires a large number of switch pixels as many as ~10[6]. A large-scale photonic switch is usually constructed from 1×2 and 2×2 elementary switches connected into networks/arrays of a variety of topologies, such as Cross-Bar, Benes, and path-independent loss (PI-Loss), as well as their numerous variants. Given that there are hundreds of or more elements in cascade, it is important to achieve high-performance elementary switches with e.g., low excess losses and crosstalk across a broad photonic bandwidth. Otherwise, even small excess loss and crosstalk (especially coherent crosstalk) from each elementary switch will accumulate rapidly as the switch networks/arrays scale up, leading to unacceptable degradation of signal integrity[16, 17]. Although photonic switches with an ultrawide wavelength-band of hundreds of nanometers are highly desired for applications such as multiband wavelength-division-multiplexing (WDM) and digital Fourier transform spectroscopy[9], they have been less reported and remain a major challenge. Also, the wavelength insensitivity often implies excellent fabrication tolerance, which is very important for realizing commercial large-scale photonic switches. Furthermore, power/energy consumption and footprints of the elementary switches are also potential limiting factors for further scaling up PICs. Finally, the overall robustness, including large tolerance for fabrication and electrical drive, excellent repeatability and stability, and long durability are all highly desired features for elementary switches. In summary, the continuous push for high-performance elementary switches with low excess loss, low crosstalk, large bandwidth, low power/energy consumption, compact footprint, and overall robustness, is one of the most crucial premises for sustainable PIC scaling as desired for many applications.

As for the operation principles, most elementary switches are realized with Mach-Zehnder interferometers[18-22] (MZIs) or photonic resonators/cavities[7, 23-25], for which two-beam or multi-beam interference is manipulated by introducing perturbative phase shift electro-optically or thermo-optically. Recently 32×32 silicon photonic switches with decent performances have been demonstrated using electro-optic (EO) MZIs[26] of Benes topology and thermo-optic (TO) MZIs[27] of PI-Loss topology. Moreover, 128×128 Benes switch array using EO MZIs is demonstrated as well, however, without characterization of the performance[28]. Since the change of the refractive index due to the EO and TO effects is usually low (i.e., ~$10^{-4}$ or ~$10^{-3}$)[29, 30], fundamental trade-offs among the footprint, the excess loss, the power consumption, and the bandwidth are inevitable. Specifically, silicon EO phase-shifters are generally millimeter-scale and lossy due to free carrier absorption[31] (FCA). Although silicon TO phase-shifters can be as short as <100 μm and almost lossless, continuous thermal power supply of tens of mW is required[29]. Alternatively, resonators/cavities can be employed to break the trade-off among the footprint, the excess loss and the power consumption, but at the price of narrow-band operations. Furthermore, due to the analog/continuous nature of the EO/TO phase shifts[32], highly precise electrical drive is often required for digital switching. These phase shifters are also highly susceptible to the fabrication variations. Therefore, it is usually required to introduce sophisticated device characterizations, feedback control and stabilization schemes[33], which becomes almost infeasible for large-scale switches with more than ~100 input/output ports. More fundamentally, phase-shifting is intrinsically a wavelength-dependent physical process[30, 34], prohibiting ultrawide operation bandwidths across hundreds of nanometers.

Another emerging type of elementary switches relies on the manipulation of mode coupling. This type of switching mechanism is well suited for micro/nano-electromechanical-system (MEMS/NEMS) actuation and transduction[35-37]. Generally speaking, for almost all MEMS photonic devices, the basic idea is to mechanically move a certain part of the structure so that the light path or phase is manipulated. For example, previous MEMS photonic switches were often realized by introducing some mechanical displacement/deformation to make the waveguide structures be coupled or decoupled[38-42]. For such type of switches, the electrostatic actuation consumes near-zero power in steady-states. MEMS switches based on directional couplers (DCs) has been demonstrated with low excess loss and low crosstalk[38, 41, 43]. However, it is not suitable for large-scale arrays in practical applications due to the bandwidth limitation of <30 nm and complex calibration caused by sensitivity to fabrication errors. Particularly, electrostatically-actuated vertical adiabatic couplers[44] (VACs) were demonstrated with low excess losses of 0.7 dB and high extinction ratio of 70 dB at the wavelength of 1500 nm. The demonstrated MEMS silicon photonic switch features a minimum footprint of 110 µm×110 µm, a drive voltage of 65 V, low switching energy consumption of tens of pJ, switching speed of µs-scale, ultrawide bandwidth of hundreds of nanometers, and excellent fabrication tolerance. Such high-performance elementary switches make it possible to realize unprecedented 240×240 and 128×128 Cross-Bar switches for interconnect[44] and Lidar applications[45], respectively. However, such VAC MEMS switches are essentially capable of only 1×2 switching operation and incapable of 2×2 switching operation[44], hence they are compatible with Cross-Bar topology only and inapplicable in many other topologies (such as Benes and PI-Loss). Due to this topological limitation, 57,600 ($240^2$) switch cells are required for the 240×240 array mentioned above, which is 30 times of a 256×256 Benes switch array with 1,920 cells. Such enormous numbers of switch cells usually significantly complicate the packaging and control of the switch array. A special approach of utilizing the hysteresis characteristics of VAC switches enables a row/column addressing scheme[46] to effectively reduce the pad scale from $N^2$ to 2N, while this approach only allows the switch cells to be switched one-after-one, as discussed in [46]. Moreover, because the Cross-Bar topology requires all the switch cells work properly to ensure proper function, such enormous numbers of switch cells are still undesired because they may compromise the yield and increase the risk of array damage. Also, the VAC design requires an additional polysilicon layer, which is incompatible with most standard silicon photonic foundry processes currently accessible[44]. Accordingly, 2×2 elementary switches based on horizontal adiabatic couplers (HACs) have been demonstrated for 8×8 Cross-Bar switches[47], where only a single silicon layer on a standard silicon-on-insulator (SOI) wafer is needed and thus low-cost standard foundries can be used. Unfortunately, the measured crosstalk in the ON state is still as high as ~−10 dB across a bandwidth of ~110 nm and the switching speed is as slow as ~36.7 μs.

Here we propose a MEMS elementary switch on standard SOI by manipulating the mode propagation, as shown in Fig. 1a, b, which is very different from the scheme of manipulating the mode interference or mode coupling. Specifically, the proposed 2×2 silicon photonic MEMS switch consists of a waveguide crossing that is split into two halves, each of which alone functions as a 90° multimode waveguide reflector based on total internal reflection (TIR). When the two halves split into two waveguide reflectors, the input light turns 90° into the adjacent waveguide on the same side, and the switch is in the OFF state, shown in Fig. 1a. On the other hand, when the two halves combine into one waveguide crossing, the input light propagates straightly and the switch is in its ON state, shown in Fig. 1b. For the present silicon photonic MEMS switch based on a split waveguide crossing (SWX), one half is stationary, while the other half is movable, actuated by a compact MEMS electrostatic comb, as shown in Fig. 1c. Such an elegant and unique structure design can drastically modify the mode propagation with an ultra-compact reconfigurable photonic structure, particularly implementing a 2×2 photonic switch (instead of the 1×2 type). Compared with the work on single-pole single-throw (SPST)[48] and the simulated single-pole double-throw (SPDT)[49] switches based on movable waveguides, the present 2×2 SWX switch features distinct operation principles and superior feasibility, performance, and scalability. Compared with the 1×2 VAC switch[44], the present 2×2 SWX switch proposed here is ~10 times more compact (for the area of the photonic structure), two-orders more energy-efficient (See Supplementary Section 8), and fully applicable in versatile topologies, with both intricate photonic and MEMS structures fabricated within a single layer of silicon on a standard SOI wafer (See Supplementary Section 1). Here the present 2×2 SWX switch features excellent performance with low excess loss of 0.1–0.52 (0.1–0.47) dB and low crosstalk of <–37.1 (–22.5) dB over an ultrawide bandwidth of 1400–1700 nm for the OFF (ON) states in simulation, while in experiment, excess loss of 0.12–0.4 (0.54–0.76) dB and crosstalk of <–44 (–24.1) dB over the bandwidth of 1420–1600 nm (limited by the input/output grating couplers) for the OFF (ON) states have been measured. The switch features a total footprint of 95 μm×100 μm (23 μm × 23 μm for the SWX alone), acceptably low drive voltage threshold of ~20 V, maximum drive voltage rating beyond ~24 V, near-zero steady-state power consumption, low switching energy of sub-pJ, switching speed of μs-scale, and durability beyond $10^9$ switching cycles (no observable performance degradation). Note that any voltage between the threshold and the maximum rating can drive the switch digitally from the OFF state to the ON state with identical performances. Such a large tolerance warrants robust drive schemes for large-scale switch networks/arrays. Note that the crosstalk in the OFF state can be much lower by splitting the two halves of the SWX more, at the price of higher drive voltage. With such high-performance 2×2 elementary switches, a 64×64 silicon photonic switch using Benes topology has been fabricated (the largest Benes switch array based on photonic MEMS reported to date) and characterized to validate the capability of the SWX switches to further scale up for practical applications such as photonic interconnect/routing, Lidar and spectroscopy, photonic computing, as well as microwave photonics.

**Results**

**Device design.** The proposed 2×2 silicon photonic MEMS switch operates with a unique SWX, which can be reconfigured electrostatically, as shown in Fig. 1. The present SWX consists of two halves configured initially with a sufficiently large air gap between them, hence they serve as compact waveguide corner-bends enabling total internal reflection (TIR). In this case the switch is in its OFF state and the input light turns 90° into the output waveguide on the same side, as shown in Fig. 1a. In order to guarantee low crosstalk in the OFF state, initially the air gap should be much larger than the Goos-Hanchen displacement, which is typically hundreds of nanometers for the telecom wavelength-band. On the other hand, when the two halves of the SWX are reconfigured electrostatically to be with a near-zero gap in-between, the structure effectively becomes a near-perfect waveguide crossing based on self-imaging, in which case the switch works in the ON state and the input light propagates straightly into the output waveguide on the other side, as shown in Fig. 1b.

To implement the proposed 2×2 silicon photonic MEMS switch, we have devised intricate and coordinated designs of photonic and MEMS structures within a single layer of silicon on standard SOI wafers, as shown in Fig. 1c. Both halves of the SWX are suspended and connected with input/output waveguides. One half is stationary, fixed by the rigid triangle formed by the straight input/output waveguides. Meanwhile, the other half is movable due to the flexibility of the meandering input/output waveguides when actuated in the $y$ direction by a compact MEMS electrostatic comb via a shuttle beam. The two halves are also connected with a pair of aligners to minimize their misalignment in the $z$ and $x$ directions (See Supplementary Section 3). Otherwise, misalignment often occurs due to the residual stress in the SOI wafer and/or asymmetric fabrication variations in the suspended silicon structures, and may lead to significant performance degradation for the ON state, when the two halves of the SWX are supposed to engage. The shuttle beam is perforated to achieve better rigidity with less mass, hence less inertia and faster switching response (See Supplementary Section 5). The shuttle is suspended from two pairs of folded springs, which are flexible in the $y$ direction but are rigid in the $x$ direction to minimize the undesired lateral displacement (See Supplementary Section 3). Between these two pairs of folded springs, there are a pair of movable electrostatic combs fixed to the shuttle beam. When a drive voltage is applied, the movable comb pair is attracted toward the stationary comb pair by the electrostatic force in the +$y$ direction, thereby engaging the SWX switch, which is now in its ON state. Meanwhile, when the drive voltage is removed, the movable comb pair is pulled back to its initial position by the restoring force (in the −$y$ direction) of the folded springs, thereby disengaging the SWX switch, which is now in its OFF state. Here the drive voltage is only applied on the stationary comb, while all other silicon structures (especially the suspended structures) are readily connected in the silicon layer and always grounded. In such a design, neither the mechanical structures nor their movement compromises the photonic performance of the rigid SWX or disrupt the input/output photonic waveguides.

It should be noted that the two halves of the SWX should never directly contact with a zero gap. Otherwise, the strong van der Waals forces between them will lead to destructive stiction problems, hence switch failure (See Supplementary Section 2). Therefore, a small bump is incorporated on both sides of the movable reflecting facet as a mechanical stopper, which can prevent large-area contact for the ON state. As a result, the contact area between the stopper and the stationary reflecting facet is small enough to prevent stiction. On the other hand, for the ON state, it is desired to minimize the gap between the two SWX halves to achieve optimal performances, which imposes stringent requirements on the fabrication precision of the stoppers. Furthermore, one should note that the sidewalls of the etched silicon layer are not perfectly vertical, which prevents the two SWX halves from engaging properly. Instead, it often results in an unacceptably large wedge-shaped gap (See Supplementary Section 2). To address these pragmatic issues, here we introduce refractive-index engineering by incorporating subwavelength-tooth (SWT) structures on both reflecting facets, as shown in Fig. 1d–f. With such a unique structural design, the two SWX halves can properly engage in the ON state because their SWT facets are matched well even when a nm-scale gap is introduced between them as desired to prevent stiction. In contrast, for the OFF state, the SWT facets are well equivalent to a thin layer with a gradient refractive index profile, still guaranteeing TIR for the incident light.

The photonic structures are designed optimally with three-dimensional (3D) finite-difference time-domain (FDTD) simulations. For the designed SWX, we choose the waveguide width $w$ = 2.68 µm and the total length $L$ = 19.54 µm. Taking the fabrication capability and the switch performance into consideration, the SWT design is finalized with a period of 290 nm and a depth of 300 nm. Note that the condition of subwavelength operation might not be satisfied as the wavelength decreases. Therefore, the SWT introduces performance degradation in short wavelengths, introducing the limitation to the ultra-broad bandwidth. To obtain low crosstalk in both states, the air gap in the OFF state is designed to be 900 nm while the height of the mechanical stoppers (corresponding to the air-gap width in the ON state) is designed to be 15 nm (See Supplementary Section 2). Figure 2a shows the simulated light propagation in the designed SWX in the OFF state, where the incident light is reflected to the output port on the same side by the SWT facet. The corresponding transmission spectra are shown in Fig. 2b, featuring an excess loss of 0.1–0.52 dB and crosstalk <–37.1 dB over an ultrawide bandwidth of 1400–1700 nm. The self-imaging imperfection as well as the ignorance of Goos-Hanchen displacement contributes to the simulated excess loss in OFF state, which shows slight wavelength dependence and can be further reduced by improving the SWX design. For the SWX operating in the ON state with a tiny gap of 15 nm induced by the mechanical stopper, the simulated light propagation is shown in Fig. 2c, where the incident light propagates straightly across the waveguide crossing area with a low excess loss and a high extinction ratio. Figure 2d shows the corresponding transmission spectra, featuring an excess loss of 0.1–0.47 dB and crosstalk of <–22.5 dB over the same bandwidth. The self-imaging imperfection caused by the mechanical stoppers contributes to the simulated excess loss in ON state primarily, which shows slight wavelength dependence as well. Evidently, the incorporation of the SWX aligner pair and the shuttle beam does not disturb the light propagation in the SWX, as shown in Fig. 2a, c, because the optical fields are negligible at the locations where these mechanical structures are connected. In addition, the elastic deformation of the mechanical structures, especially the meandering waveguides, causes no additional excess loss as well.

Figure 2e shows the schematic diagram of the springs used here, designed with high rigidity in the $x$ direction and high flexibility in the $y$ direction (See Supplementary Section 3). Figure 2f shows the schematic diagram of the electrostatic combs. The width $w_{st}$ of the stationary comb handle is designed to be 6 µm, which is sufficiently wide to remain anchored to the buried oxide (BOX) layer during the MEMS releasing process. The width $w_{dt}$ of the movable comb handle is designed to be 3 µm, which is sufficiently narrow to be completely released, meanwhile sufficiently rigid to avoid bending due to the electrostatic and elastic force. To ensure excellent stability for the entire MEMS structures in the $x$ direction, the electrostatic force generated in the *x* direction should be minimized by meticulously designed comb fingers. Here the overlap length ($l_c - a_c$) and the gap width $d_c$ are designed to be 200 nm and 300nm, respectively (See Supplementary Section 4 and 6). Figure 2g shows the electrical and mechanical simulation results using finite element method (FEM). The red line shows the linear relationship between the driving force and the displacement. The elastic deformation of the mechanical structures follows Hooke's law, with a spring constant of ~0.33 N/m. The blue curve shows the quadratic relationship between the driving force generated by the electrostatic combs and the applied voltage. According to the electrical and mechanical simulation results, the threshold voltage is ~22 V for the designed SWX switch with an initially 900-nm-wide gap in the OFF state. Finally, the entire MEMS structure design has been verified with the FEM simulation of the ON state shown in Fig. 2h, where the maximum tensile and compressive principal strain of ~±3×10$^{-3}$ occurs at the SWX aligners, well below the damage threshold of silicon (See Supplementary Section 3). To analyze the frequency response of the device, the fundamental resonance frequency of the mechanical structures is estimated to be 0.14 MHz as a simple harmonic oscillator. Such a resonance frequency indicates that the highest switching frequency is up to ~100 kHz, enabling μs-scale switching speed (See Supplementary Section 5). Resistance of the device against mechanical shocks are estimated in theory as well (See Supplementary Section 7). The results show that accelerations up to 100*g* (*g* is the gravitational acceleration) only cause minor disturbances to the structures.

**Experimental results.** The SWX switch device was fabricated with electron-beam lithography (EBL) process in which proximity effect correction (PEC) is involved. Figure 3a shows a scanning electron microscope (SEM) image of the device, where the stationary and the movable SWX halves are highlighted with red and blue, respectively. All structures are suspended except the stationary electrostatic comb and the anchors. The SWX itself has an ultra-compact footprint of 23 μm×23 μm, while the total footprint including the MEMS actuator is 95 μm×100 μm. As shown by the close-up view of the SWT structures and mechanical stoppers in Fig. 3b, the period and depth of the SWT were measured to be 290 nm and 285 nm approximately, while the height of the mechanical stopper was measured to be about 15 nm. The movable half-SWX is 900 nm away from the stationary one in the OFF state as designed. When the switch is turned on, the movable half-SWX is pushed toward the stationary one by the MEMS actuator and the gap between the two halves is precisely defined to be 15 nm by the mechanical stoppers.

For precise measurement of the excess loss, ten cascaded elementary switches were measured and averaged. Figure 3c shows the measured transmission spectra $T_{11}$ and $T_{12}$ at the two output ports in the OFF state, exhibiting a low excess loss of 0.12–0.4 dB for $T_{11}$ and a low cross talk of <–44dB for $T_{12}$ in the measured wavelength range of 1420–1600 nm (further broadband measurement is limited by the efficiency of the grating couplers used here). Figure 3d shows the measured transmission spectra $T_{11}$ and $T_{12}$ at the two output ports in the ON state, where the excess loss of $T_{12}$ is 0.54–0.76 dB and the crosstalk of $T_{11}$ is <–24.1 dB in the wavelength range of 1420–1600 nm. For the OFF state, the excess loss agrees well with the simulated results, while for the ON state the excess loss is slightly higher due to fabrication imperfection of the SWX and the SWT. The measured crosstalk in the ON state agrees with the simulated value (<–22.5 dB), indicating that the SWX is engaged tightly with a gap of <15 nm and the sidewall inclination is small.

To further characterize the steady-state switching performance, transmission spectra $T_{12}$ with different drive voltages were measured and shown in Fig. 3e. The threshold voltage was measured to be 20 V, which agrees well with the simulation result of 22 V. The switch remains in the ON state without damage with a drive voltage up to 30 V, indicating a maximum drive voltage rating beyond 30V. Because the actual maximum rating cannot be obtained without damaging the device, we refrained from such measurements. Note that any voltage between the threshold and the maximum rating can drive the switch digitally into its ON state with identical performances, as shown in the spectra measured with the drive voltage of 20–30 V in Fig. 3e. Such a large tolerance warrants robust drive schemes for large-scale switch arrays. The temporal response of the switch was measured by applying a square-wave pulse, as shown in Fig. 3f. The measured ON and OFF switching time (to 90% power and to 10% power) are respectively ~3.5 μs and ~1.2 μs. Note that a lower spring constant is preferred for achieving a lower threshold voltage, which however compromises the switching speed. For the ON switching, the movable half-SWX needs to travel ~600 nm forward (+$y$ direction) from the initial OFF position to achieve sufficient optical power redirection from one half-SWX to the other, while for the OFF switching, the movable half-SWX only needs to travel ~300 nm backward (–$y$ direction) from the ON position to stop the optical power redirection. Therefore, the OFF switching is faster than the ON switching. In order to test the durability of the SWX switch, we operated the switch for over one billion cycles with a square-wave voltage at 20 kHz. The measured transmission $T_{12}$ at 1550 nm in the OFF and ON states are shown in Fig. 3g. Here the cross marks represent the measured transmission every two hours, exhibiting no performance degradation during the 14-hour test (See Supplementary Section 10). Finally, the SWX switch was held in the ON state over one hour without stiction (at the mechanical stoppers) or performance degradation.

Based on the 2×2 SWX switch described above, a 64×64 switch array using Benes topology was designed and fabricated by manual EBL process. To our best knowledge, this is the largest Benes switch array based on photonic MEMS reported to date. The switch array consisting of 352 switch cells has a footprint of 10 mm×5.3 mm, as shown in Fig. 4a. Figure Fig. 4b, c, shows the close-up view of the switch cell and the waveguide connection shuffle. As shown in Fig. 4d, waveguides in the connection shuffles are broadened to reduce propagation loss and bent in Euler curve shape[50] to reduce footprint as well as suppress high-order modes (See Supplementary Section 12). Meanwhile, varied-width multimode interference (MMI) waveguide crossings[51] are used in the connection shuffle to achieve low-loss and low-crosstalk propagation (See Supplementary Section 12). Grating couplers are used to access the input/output ports. The fabricated switch array was characterized with a pair of 66-channel fiber arrays and a pair of electrical probes (See Supplementary Section 13). The switch array in the initial all-OFF state (all 352 switch cells are OFF) was characterized by coupling light into the 64 input ports sequentially, and the measured transmission spectra $T_{i,j}$ ($i, j$ = 1, …, 64) are shown in Fig. 4e. The maximum crosstalk of the leakage paths ($T_{i,j}$ with $i \neq j$) is 35 dB lower than the minimum transmission of the optical paths ($T_{i,j}$ with $i = j$ = 1, …, 64) over the C band, exhibiting high crosstalk suppression. The excess losses of the 64 optical paths at the center wavelength of 1550 nm are shown in Fig. 4f by the blue rings. Notably, among the 64 optical paths, there are 6 paths ($i = j$ = 5, 21, 22, 48, 54, 55) exhibiting conspicuously higher losses, as denoted by the red crosses in Fig. 4f. These elevated losses are attributed to specific fabrication defects that were inadvertently introduced during the manual EBL process (See Supplementary Section 12). To avoid confusion between defective optical paths and leakage paths, transmission spectra of these six defective optical paths are excluded in Fig. 4e. The excess losses of the 58 intact optical paths vary from 8.0 dB to 12.6 dB at the wavelength of 1550 nm. Meanwhile, we estimate the excess losses to be 5.2–11.7 dB, as the red rings show in Fig. 4f, according to our systematic experimental characterizations of the excess/propagation loss of the switch cells, as well as the waveguides of different widths, varied-width MMI waveguide crossings and Euler bends in the connection shuffles of the Benes array (See Supplementary Section 11). Some discrepancies between the measured and estimated excess losses of the 58 intact optical paths are attributed to some slight fabrication defects, variations of the grating couplers (See Supplementary Section 11), inter-channel variations of the fiber arrays and the random particulate contamination directly on top of the silicon structures. Since the switch cells were electrically addressed by a pair of probes manually in the experiment, further measurement was limited to 32 specific states with single-ON switching (all switch cells are OFF except the selected one). Here the 32 switch cells in the center stage of the Benes array were switched ON sequentially, and the measured ON/OFF transmission spectra of the corresponding output ports are shown in Fig. 4g, exhibiting extinction ratios > 38 dB over the C band, with the results from four defective paths excluded (See Supplementary Section 12).

**Discussions**

**Immunity to thermomechanical displacement noise.** Thermomechanical displacement noise is present in any mechanical structures at finite temperature, and is most pronounced for nanomechanical structures near eigenmode (resonance) frequencies[52-54]. Nevertheless, the effect of such noise on the SWX switch performance is negligible. For the ON state, the two SWX halves engage, such that the air gap in-between is mechanically fixed, even if the entire suspended structures are undergoing random thermomechanical motion. For the OFF state, the two SWX halves disengage, such that the air gap in-between is orders of magnitude larger than the thermomechanical displacement of the two SWX halves. Therefore, in neither of the two states, the displacement noise can be transduced onto the photonic signals. The same arguments apply to displacement noise of other origins.

**Topology and scalability.** Note that our 2×2 elementary switch has two input ports and two output ports, which is totally different from and much more versatile than those 1×2 switches that are limited to the Cross-Bar topology. Utilization of the present 2×2 elementary switch enables highly flexible large-scale switch network/array designs with various topologies such as Benes, PI-Loss and Cross-Bar. Note that all the array topologies feature specific advantages and disadvantages, therefore, the compatibility with various topologies is beneficial, so that it is possible to make best use of the advantages and avoids the disadvantages by choosing different topologies flexibly for different applications. For an N×N switch array with the Cross-Bar topology, signals transmitted from any input port to any output port pass only one elementary switch that is in the ON state. Accordingly, if the elementary switches work with low losses in the OFF state (which can be achieved in most cases), the total loss can be low when low-loss waveguide crossings are used, which is a unique advantage. However, for the Cross-Bar topology, $N^2$ elementary switches are required for an N×N switch array. The number of elementary switches increases quickly as the port counts scale up, which may compromise the yield and increase the risk of array damage. Furthermore, for switches without hysteresis characteristics, hence the unique row/column addressing scheme of the VAC switch[46] is not applicable, switch arrays with such enormous switch numbers may require challenging and cumbersome packaging and a sophisticated control system. In contrast, when using the Benes topology, one only needs $N(\log_2 N - 0.5)$ elementary switches for an N×N switch array. In this case, the total number of elementary switches can be minimized, which may significantly increase the yield and simplifies the packaging and the control system. In addition, for Benes topology, there are only $2\log_2 N - 1$ elementary switches (ON or OFF) in cascade for any optical paths[17], resulting in acceptable accumulated excess loss (compared with that N switch cells in cascade are needed for Cross-Bar topology). In light of the excellent scalability of the Benes topology, it is our on-going effort to develop a large-scale N×N switch with e.g., N = 128 or more to meet the demands in various applications, especially photonic interconnect/routing and photonic computing.

**Interfacing strategies.** For MEMS switch arrays, the electrical pads are usually spread out across the entire layout because both the pads and wires are evaporated on the silicon directly, which requires electrical isolations, and thus the routing space of electrical wires is greatly constrained. The interfacing of Benes switch array, which minimizes the number of pads effectively, can be readily implemented with mature technologies such as electrical feedthrough[55] and flip-chip bonding[56], both of which are compatible with standard silicon photonic foundry processes.

**Compatibility with foundry processes.** As described above, the fabricated reflecting-facets are not perfectly vertical in most cases, which leaves a wedge-shaped air gap between the engaged SWX halves. Therefore, the SWT structure is introduced to alleviate the performance degradation due to the small air gap when the SWX switch is ON. For the fabricated devices in this paper, the SWT is designed with a period of 290 nm and a depth of 300 nm to achieve high performances. Note that such SWT with a relatively high aspect ratio (depth/period) of 300/290 cannot be fabricated yet by most commercial standard silicon photonic foundries with a minimum feature size of 130 nm or 180 nm (130/180-nm processes). Fortunately, there are a few advanced foundries with the minimum feature size of 90 nm and below available, which are potentially capable of fabricating the SWT. For the mechanical stopper even with tinier size than the SWT, it is easier to be fabricated in standard silicon photonic foundries due to its low lateral aspect ratio. In this paper all the devices were fabricated with EBL for experimental verification.

Alternatively, if the fabrication can be improved to further reduce sidewall inclination, a smaller air gap can be achieved for the SWX in the ON state, such that shallower SWT can be used for the reflecting facet. For the ideal case with zero air gap, excellent switch performance can be achieved even with flat reflecting facets (without SWT), which can be fabricated by a regular foundry with 130/180-nm processes for silicon photonics. Since it is nontrivial to consistently fabricate perfectly vertical silicon sidewalls, and zero gap is not allowed to avoid stiction, a feasible compromise is to push for gap widths less than 10 nm. In this case, excellent switch performance can still be obtained in the ultrawide bandwidth of 1400-1700 nm even when the SWT depth is reduced to 200 nm (See Supplementary Section 2), requiring the sidewall tilt angle to be <~1.5° (which is possible with a well-calibrated etching process). Accordingly, such SWT can be fabricated compatibly with commercial standard silicon photonic foundries with low-cost 130-nm or even 180-nm processes.

As it is well known, both the waveguide propagation loss and the waveguide-crossing loss are critical sources for the Benes-topology array. When using commercial standard foundries, the propagation loss of the 600-nm-wide silicon photonic waveguides can be as low as 1.1 dB/cm or less, which is six times smaller than the loss of our silicon photonic waveguide fabricated with the manual EBL process. Meanwhile, as demonstrated in our previous work, the excess loss of the varied-width MMI waveguide crossings can be reduced to <10 mdB[51] with standard silicon photonic foundries. As a result, the total excess loss of the 64×64 array can be reduced to 6.3 dB in average and that of a 128×128 array can be expected as 9.1 dB (See Supplementary Section 12), such that the SWX switch will become a highly viable solution for large-scale silicon photonic switches for a broad spectrum of practical applications such as photonic interconnect/routing, Lidar and spectroscopy, photonic computing, as well as microwave photonics.

**Methods**

**Fabrication of the device.** The devices were fabricated on commercial SOI wafers (SOITEC) with a 220-nm-thick top silicon layer and a 2-μm-thick BOX layer. First, the ridge waveguides and strip waveguides were patterned with EBL, followed by the processes of 150-nm shallow etching and 220-nm full etching respectively. Second, metal electrodes (50-nm-thick chrome and 300-nm-thick gold films) were patterned with photolithography, followed by E-beam evaporation and lift-off processes. Finally, the suspended structures of the devices were released from the BOX by hydrofluoric vapor etching. The fabrication process for our device is not complicated and fully compatible with commercial standard silicon photonic foundries as well as standard silicon photonic MEMS platform.

**Measurement of the device.** Both the 2×2 elementary switch and the 64×64 Benes switch array were measured on a fiber-to-chip coupling platform (See Supplementary Section 13). An amplified spontaneous emission laser (1520–1610 nm) was used as the light source and the drive voltage signals were applied to the devices through direct-current probes. All the optical transmission spectra were measured with an optical spectrum analyzer (OSA, Yokogawa, AQ6370D). For the measurement of the elementary switch, a pair of fibers were used to couple the optical signal. The fiber-to-chip coupling loss was measured to be ~5 dB. In the ON/OFF optical transmission spectra measurement, the drive voltage was generated by a sourcemeter (Keithley 2400). The square-wave voltage for the temporal response measurement and the durability measurement was generated by a signal generator (SIGLENT SDG1032X). In the temporal response measurement, light output from the switch was measured with a photodiode (Newport 2053) and an oscilloscope (Tektronix, TBS1000C). For the measurement of the 64×64 switch array, a pair of 66-channel fiber array (spaced 80 µm, angled 8°) was used to couple the optical signal. To measure the 4096 transmission spectra of the Benes switch in its initial (all-OFF) state, a pair of commercial 1×64 optical switches were used to perform automatic measurements (See Supplementary Section 13).

**Data availability**

Data underlying the results presented in this paper are available from the authors upon request.

## Acknowledgements

This work is funded by National Science Fund for Distinguished Young Scholars (61725503), National Natural Science Foundation of China (U23B2047, 62321166651, 92150302), Leading Innovative and Entrepreneur Team Introduction Program of Zhejiang (2021R01001), Zhejiang Provincial Major Research and Development Program (2021C01199), Natural Science Foundation of Zhejiang Province (LZ22F050006), Fundamental Research Funds for the Central Universities, and Startup Foundation for Hundred-Talent Program of Zhejiang University. The authors thank the ZJU Micro-Nano Fabrication Center and the Westlake Center for Micro/Nano Fabrication and Instrumentation for the facility support.

## Author contributions

D. D. conceived the project. Y. H., Y. Sun, H. L., and D. D. designed the structures. Y. H. fabricated the devices. Y. H., and Y. L. characterized the devices. Y. H., H. L. and D. D. contributed to the data analyses. D. D., H. L., L. L., and Y. Shi managed the project. Y. H., H. L. and D. D. wrote the manuscript. All authors discussed the results and contributed to the manuscript.

## Competing interests

D. D. and Y. Sun have a granted patent (grant number: CN110658584B/US11598921, PCT application number: PCT/CN2020/076340) on the silicon photonic MEMS switches based on split waveguide crossing presented in this paper. D. D., Y. Sun, and H. L. have a granted patent (grant number: CN112305676B) on the silicon photonic MEMS switches based on split waveguide crossing presented in this paper, which is also in progress of PCT patent application (application number: PCT/CN2021/078381). D. D., Y. H., and H. L. have filed a patent application (application number: CN116299874A) on the silicon photonic MEMS switches based on split waveguide crossing presented in this paper. The remaining authors declare no competing interests.

**Figure 1**

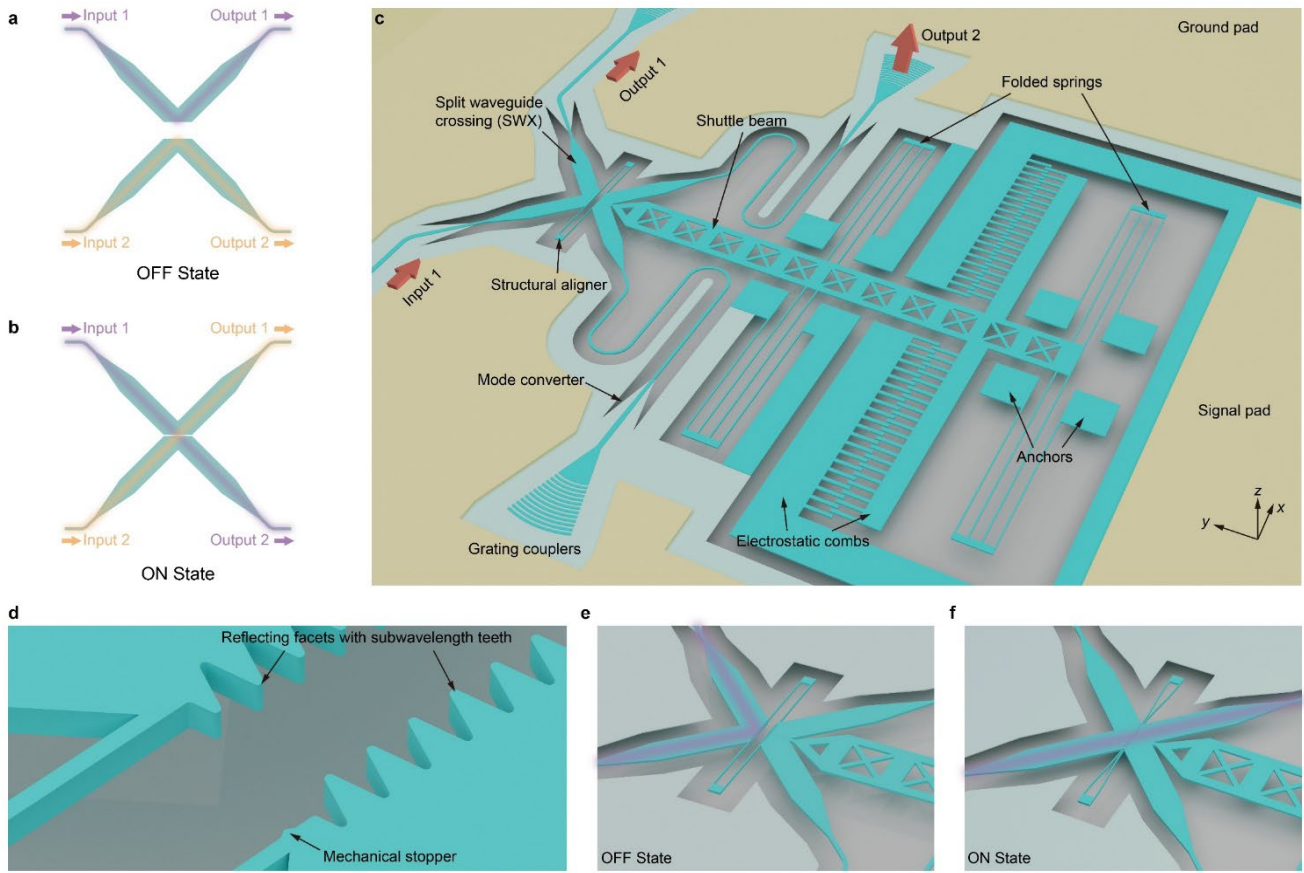

**Fig. 1 | Schematic illustrations of the switch. a, b,** The split waveguide crossing (SWX) in the OFF and ON state, respectively. **c,** Schematic diagram of the designed photonic and electromechanical structures. **d,** Close-up view of the subwavelength-teeth (SWT) and mechanical stopper. **e, f,** The SWX with SWT reflecting facets of the switch in the OFF and ON state, respectively.

**Figure 2**

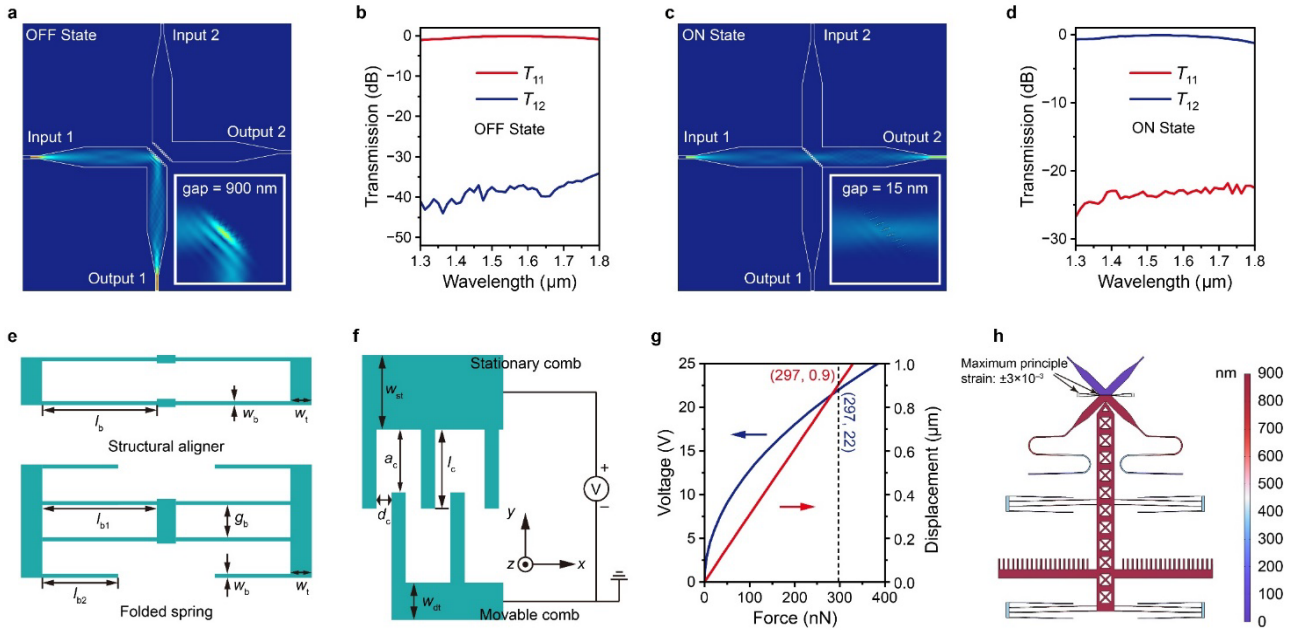

**Fig. 2 | Design of the switch based on split waveguide crossings.** Simulated light propagation diagrams with transmission spectra of the switch in the OFF state (**a, b**) and ON state (**c, d**), respectively. The inset of **a** and **c** shows the close-up view around the reflecting facets. Schematic diagram of the structural aligner, folded spring (**e**), and electrostatic comb (**f**). **g,** Simulated electrostatic force arisen by different voltage and corresponding displacement of the movable mechanical structures. **h,** Displacement of the movable mechanical structures in the ON state.

**Figure 3**

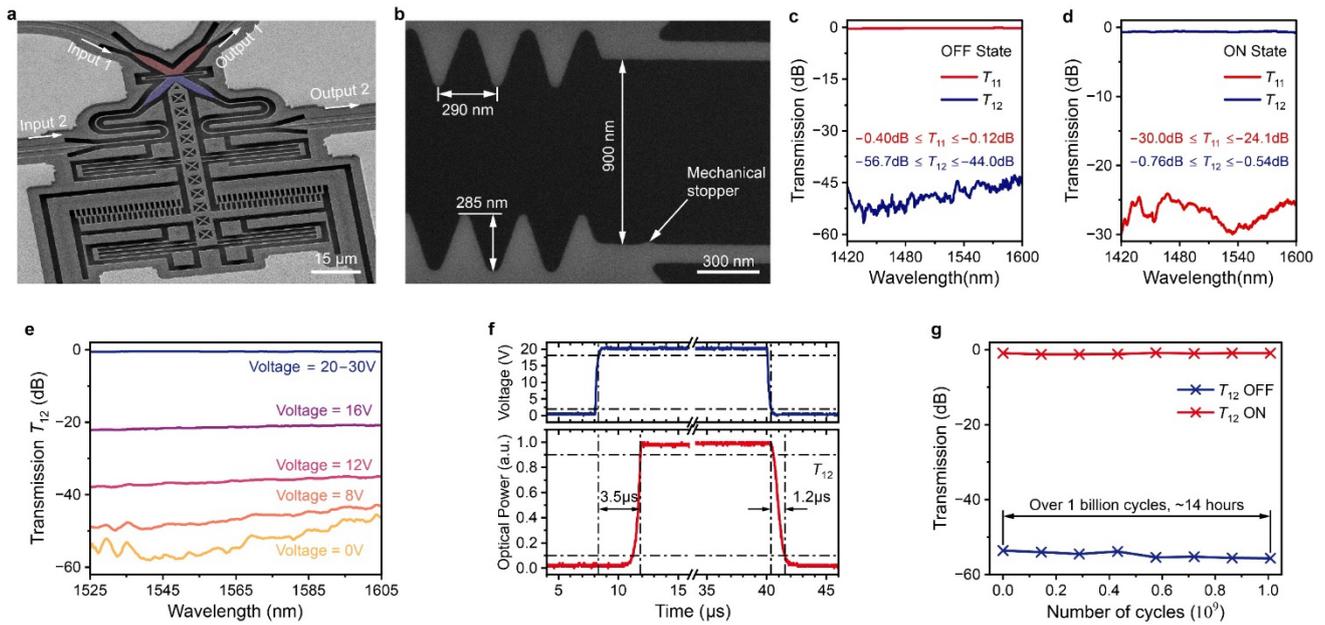

**Fig. 3 | Fabrication and measurement results of the device. a,** Perspective scanning electron microscope (SEM) image of the fabricated device. Stationary and movable halves of the split waveguide crossing are highlighted with red and blue, respectively. **b,** A close-up view of the reflecting facets and mechanical stopper. Period and depth of the subwavelength-teeth were measured. **c, d,** Transmission $T_{11}$ and $T_{12}$ in the OFF and ON state, respectively. **e,** Transmission $T_{12}$ with different actuation voltages. The threshold voltage is measured to be 20 V and the maximum drive voltage rating is beyond 30 V. **f,** Temporal response of the switch. **g,** Durability test of the switch. The switch was turned on and off for over 1 billion times and the optical transmission $T_{12}$ at 1550nm was recorded.

# Figure 4

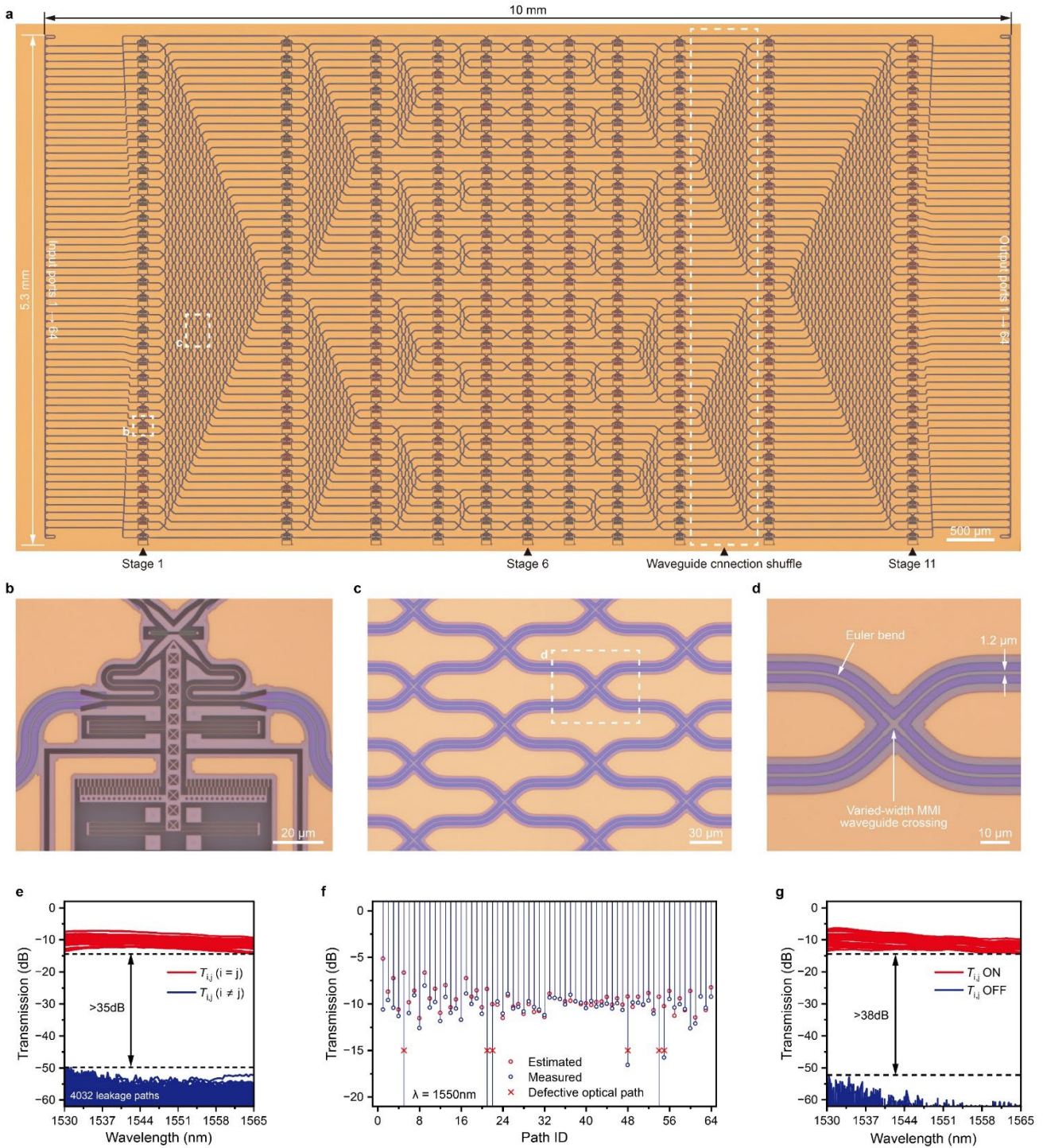

**Fig. 4 | Fabrication and measurement results of the 64×64 switch array. a,** The optical microscope image of the 64×64 switch array. **b, c, d,** Close-up view of a switch cell, connection shuffle (rotated by 90°) and varied-width multimode interference waveguide crossing, respectively. **e,** The measured transmission $T_{i,j}$ of the switch array in the all-OFF state. **f,** Estimated and measured excess loss of the 64 optical paths in the all-OFF state at 1550 nm. **g,** The measured transmission $T_{i,j}$ in the OFF/ON states of corresponding output ports with a single switch cell in the center stage turned on.

# Supplement Information: Silicon photonic MEMS switches based on split waveguide crossings

## Section 1: Reported silicon photonic elementary switches

As shown in Table 1, our device, enabled by the unique principle of manipulating the mode propagation, shows excellent performance in every aspect compared to the switches based on manipulation of mode interference or mode coupling.

• **Footprint.** When considering the photonic structure only, our 2×2 SWX switch is one of the most compact structures among various optical switches (even including 1×2 switches, as shown in Table 1). It can be seen that our switch has a photonic structure as compact as 23×23 $\mu m^2$, which is slightly larger than the smallest footprint of 6×12 $\mu m^2$ for the SPDT 1×2 switch reported in [13]. Benefitting from the compact photonic structure, our device is relatively compact as well even for the full footprint including the MEMS actuator, which can be further miniaturized to reduce the full footprint.

• **Excess loss, crosstalk and bandwidth.** Our device features low excess of 0.12–0.4 dB (0.54–0.76 dB) and low crosstalk of < –44 dB (< –24.1 dB) in OFF (ON) state over the ultra large bandwidth of 1420–1600 nm in measurement, which is the best broadband switch performance to our known. The only device with comparable simulated bandwidth is the switch based on VACs, which shows excellent OFF state excess loss and crosstalk over large bandwidth (detailed data is not shown), however, at the price of two layers of silicon and incompatibility with standard foundry process. Furthermore, when the switch based on VAC is turned to ON state, it features higher excess loss of 0.4–1.6 dB and higher crosstalk of < –17.5 dB over a narrower bandwidth of 1460–1580 nm compared to our device, showing a –20-dB bandwidth of < 60 nm. However, the SWX switch features –20-dB bandwidth of > 180 nm, which is three times as large. Consequently, the SWX switch is suitable for more applications with broadband requirements, such as spectrometers and coarse wavelength division multiplexing (CWDM) systems.

• **Voltage.** Our device features digital switch function with a threshold voltage of 20 V, which is relatively low compared to other MEMS digital switches.

• **Power/Energy consumption.** Our device features ultra-low reconfiguration energy consumption of 0.54 pJ and standby power consumption of less than 4 nW due to the tiny capacity and leakage current, which is far lower than EO / TO switches necessitating continuous power consumption of mW and two orders of magnitude lower than the VAC switch.

• **Speed.** Our device features ON/OFF switching time of 3.5 μs /1.2 μs, which is faster than most of the proposed MEMS switches. The SPDT switch achieves faster switching (0.82 μs/ 0.72 μs) at the price of far higher crosstalk (<–23 dB). The VAC switch achieves faster switching (0.4 μs/ 0.3 μs) at the price of far higher drive voltage (60 V) and reconfiguration energy consumption (42 pJ).

• **Durability.** The VAC switch demonstrates durability of > 10 billion cycles, which is encouraging as they provide ample evidence for the durability of photonic MEMS structures. Since we only presented results for over 1 billion cycles and no observable performance degradation happened in our tests, it is possible to have over 10 billion testing by prolonging the time for testing.

• **Scalability.** The 64×64 presented in this work is the largest Benes switch array based on photonic MEMS. Enabled by the unique SWX, our switch features 2×2 switch function, which makes it compatible with all mainstream array topologies. We choose Benes topology, which is forbidden for the 1×2 switches, due to its characterization of minimizing the switch counts in large scale arrays. As shown in Table 1, the 64×64 Cross-Bar array needs $64^2 = 4,096$ cells, which is more than 10 times of the 352 cells necessitated by our 64×64 Benes array. If two 256×256 switch arrays are implemented using Cross-Bar and Benes topology, they would require 65,536 and 1,920 switch cells respectively. The former is more than 30 times larger than the latter, which may compromise the yield and increase the risk of array damage.

• **Fabrication Simplicity.** The fabrication of the SWX switch with a single layer of silicon is much simpler compared to the VAC switch which demonstrates performance competitive with our device but requires an extra silicon layer.

• **Potential applications.** Enabled by the unique SWX structure, the photonic switch proposed in this paper features excellent performance, 2×2 functionality and ultrabroad bandwidth. Consequently, it can handle not only the applications that have been demonstrated by 1×2 MEMS switches in the literatures, but also more applications in which ultrabroad bandwidth and 2×2 functionality are required while conventional switches are infeasible. Firstly, the 2×2 SWX switch can be used as 1×2 switch to implement Cross-Bar switch arrays for photonic interconnect/routing (especially for AI large models) and Lidar applications that have been demonstrated with 1×2 MEMS VAC switches. Secondly, the ultrabroad bandwidth of the SWX switch enables more applications with broadband requirements, such as spectrometers and coarse wavelength division multiplexing (CWDM) systems. Thirdly, the 2×2 functionality of the SWX switch can be used for systems such as microwave photonics and Fourier spectrometers where delaylines with different delay time are reconfigured by photonic switches.

# Table 1. Summary of the reported silicon photonic elementary switches

| Types | Mechanism | Ref. | Footprint (μm × μm) (Photonic structures only) | Full footprint (μm × μm) | Excess Loss (dB) | | Crosstalk (dB) | | Bandwidth (nm) | | Voltage (V) | Power/Energy Consumption | Speed | Durability | Type | Array scale, array topology, switch counts | Standard Fabrication |
|---|---|---|---|---|---|---|---|---|---|---|---|---|---|---|---|---|---|
| | | | | | Simulated | Measured | Simulated | Measured | Simulated | Measured | | | | | | | |
| Mode Interference | Electro-optic | 1 | 400×50 | 400×50 | 0.8 | 2.9 | −20 | −18 | 110 | 110 | 1 | 3mW | 4ns / 0.65ns | − | 2×2 Analog MZI | No array | Yes |
| | Thermo-optic | 2 | 110×30 | 110×30 | 0.1 | 1 | −20 | −20 | 148 | 140 | − | − | − | − | 2×2 Analog MZI | No array | Yes |
| | | 3 | 200×30 | 200×30 | 0.2 | 0.5 | −32 | −25 | 35 | 35 | − | 26mW | − | − | 2×2 Analog MZI | No array | Yes |
| Mode Coupling | MEMS | 4 | 32×10 | 60×65 | − | 1 | − | −6.9 | − | − | 25.3 | − | ~50 μs | − | 2×2 Analog DC | 2×6 − 5 | Yes |
| | | 5 | 50×10 | 110×110 | − | 0.3 | − | −17 | 35 | | 35 | − | − | − | 2×2 Analog DC | No array | Yes |
| | | 6 | 80×80 | 160×160 | − | 2.47 | −25 | −26 | − | 13 | 14 | − | 2.5 μs / 3.8 μs | − | 1×2 Analog DC | 50×50 Cross-Bar 2,500 | Yes |
| | | 7 | 100×100 | 170×170 | − | 0.57 | − | −60 | − | 31.5 | 9.6 | − | 9.8 μs / 4.8 μs | − | 1×2 Analog DC | 4×20 Cross-Bar 80 switch cells | Yes |
| | | 8 | 90×90 | 166×166 | − | 0.5 | −52 | −51 | − | 29 | 10 | − | 50 μs / − | − | 1×2 Analog DC | 32×32 Cross-Bar 1,024 | Yes |
| | | 9 | 80×80 | 110×110 | − | 0.7 | − | −70 | − | − | 34.5 | − | − | >10^10 | 1×2 Digital VAC | 50×50 Cross-Bar 2,500 | No |
| | | 10 | 80×80 | 110×110 | 0.01 | 0.47 | −45 | −60 | 300 | 60 | 60 | − | 0.91 μs/ 0.28 μs | >10^10 | 1×2 Digital VAC | 64×64 Cross-Bar 4,096 | No |
| | | 11 | 80×80 | 110×110 | − | 0.7 | − | −70 | 300 | − | 65 | 42 pJ | 0.4 μs / 0.3 μs | − | 1×2 Digital VAC | 240×240 Cross-Bar 57,600 | No |
| | | 12 | 70×70 | 125×125 | − | 0.9 | − | −37 | − | − | 18 | − | 36.7 μs/ 21.4 μs | − | 1×2 Digital HAC | 8×8 Cross-Bar 64 | Yes |
| | | 13 | 6×12 | 65×62 | − | − | − | −23 | 100 | 25 | 22 | − | 0.82 μs/ 0.73 μs | − | SPDT Digital HAC | No array | Yes |
| | | 14 | 146×70 | 146×96 | − | 1 | − | −26 | − | 35 | 3.75 | − | − | − | 1×2 Analog DC | No array | Yes |
| Mode Propagation | MEMS | This work | 23×23 | 100×90 | 0.1 | 0.54 | −37 | −49 | 300 | 180 | 20 | 0.54 pJ | 3.5 μs / 1.2 μs | >10^9 | 2×2 Digital SWX | 64×64 Benes 352 | Yes |

## Section 2 : Design of the proposed split waveguide crossing (SWX)

The finite-difference time-domain (FDTD) method is used for simulating the light propagation in the photonic waveguide structures considered here. The waveguide crossing is designed by using the self-image in a multimode interference section. Figure S1a shows the schematic diagram of a regular waveguide crossing without splitting. In this design, the lengths of the taper and the multimode waveguides are chosen as $L_1$ = 6 μm and $L_2$ = 19.54 μm, respectively, while the widths of the singlemode and the multimode section are chosen as $w_1$ = 0.4 μm and $w_2$ = 2.68 μm, respectively. Figure S1b shows the theoretical transmissions at the through port and the cross port. It can be seen that the excess loss is 0.05–1 dB while the crosstalk is < –32.2 dB. –in the wavelength range of 1300–1800 nm. Note that the excess loss is as low as 0.05–0.42 dB while the crosstalk is < –37 dB in the wavelength range of 1400–1700 nm. Figure S1c shows the simulated light propagation in the waveguide crossing when operating at 1550nm.

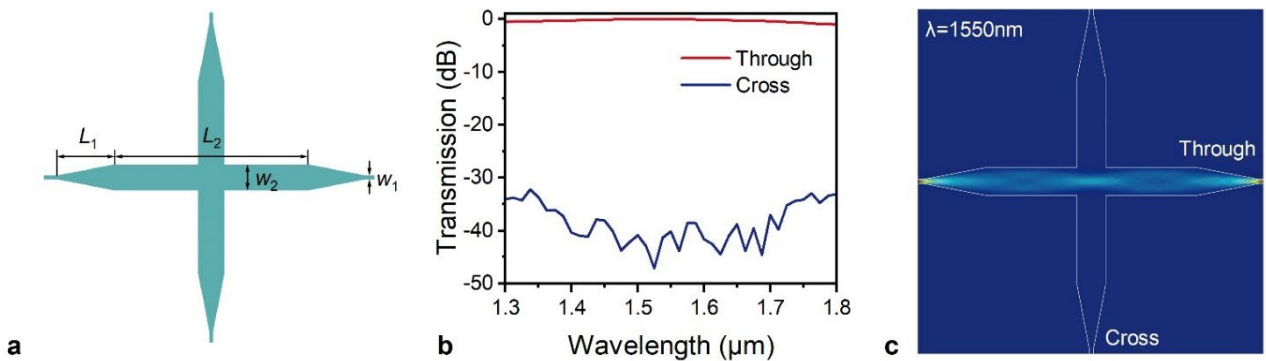

**Fig. S1 | Design of the waveguide crossing. a**, The structure. **b**, Transmissions of the through port and the cross port . **c**, Simulated light propagation when operating at 1550nm in the designed waveguide crossing.

For the proposed split waveguide crossing (SWX), there are subwavelength teeth (SWT) structures on the reflecting facets, as shown in Fig. S2a, where $p$ and $d$ are the period and depth of the SWT respectively, $g$ is the gap width between the two facets. Here the structural aligners and the mechanical stoppers are introduced, as shown in Fig. S2a. These structures were not taken into consideration for the optical design because little influence is introduced to the light propagation. Fig. S2b shows the reflection of the SWX with different periods $p$. Here the depth $d$ = 300 nm and the gap width $g$ = 1 μm (i.e., in the OFF state). It can be seen that the reflection at the short wavelength (1300 nm) is improved significantly as the SWT period $p$ decreases. This is because the subwavelength condition is better satisfied for the short wavelength when choosing a smaller period. In order to achieve high performance in a broad wavelength-band, we choose $p$ = 290 nm in our design considering the fabrication capability. With this design, the reflection loss $T_{11}$ is 0.11 – 0.52 dB in an ultra-wide wavelength-band of 1400-1700 nm. The loss is even as low as 0.11 – 0.29 dB in the wavelength range of 1450 – 1650 nm. We also simulate the SWT with different depths $d$, as shown in Fig. S2c. It can be seen that the reflection loss $T_{11}$ increases slightly from 0.08 dB to 0.2 dB at the wavelength of 1550 nm as the depth $d$ increases from 0 to 600 nm.

On the other hand, note that the depth also plays a key role for determining the switch performance in the ON state. Assume that there is a 20-nm-wide gap between the two halves of the switch ($g$ = 20 nm) in order to avoid stiction when operating in the ON state. Figure S2d shows the simulated transmission $T_{12}$ as the depth $d$ increases from 0 to 400 nm. It can be seen that the transmission loss $T_{12}$ is reduced when increasing the depth $d$. Here we choose $d$ = 300 nm to balance the performances in the OFF and ON states. With this design, the losses for the OFF state $T_{11}$ and ON state $T_{12}$ are respectively 0.11–0.52 dB and 0.12–0.55 dB in the wavelength range of 1400–1700 nm. If a narrower bandwidth of 1450–1650 nm is considered, the $T_{11}$ and $T_{12}$ are 0.1 – 0.28 dB and 0.12 – 0.39 dB respectively

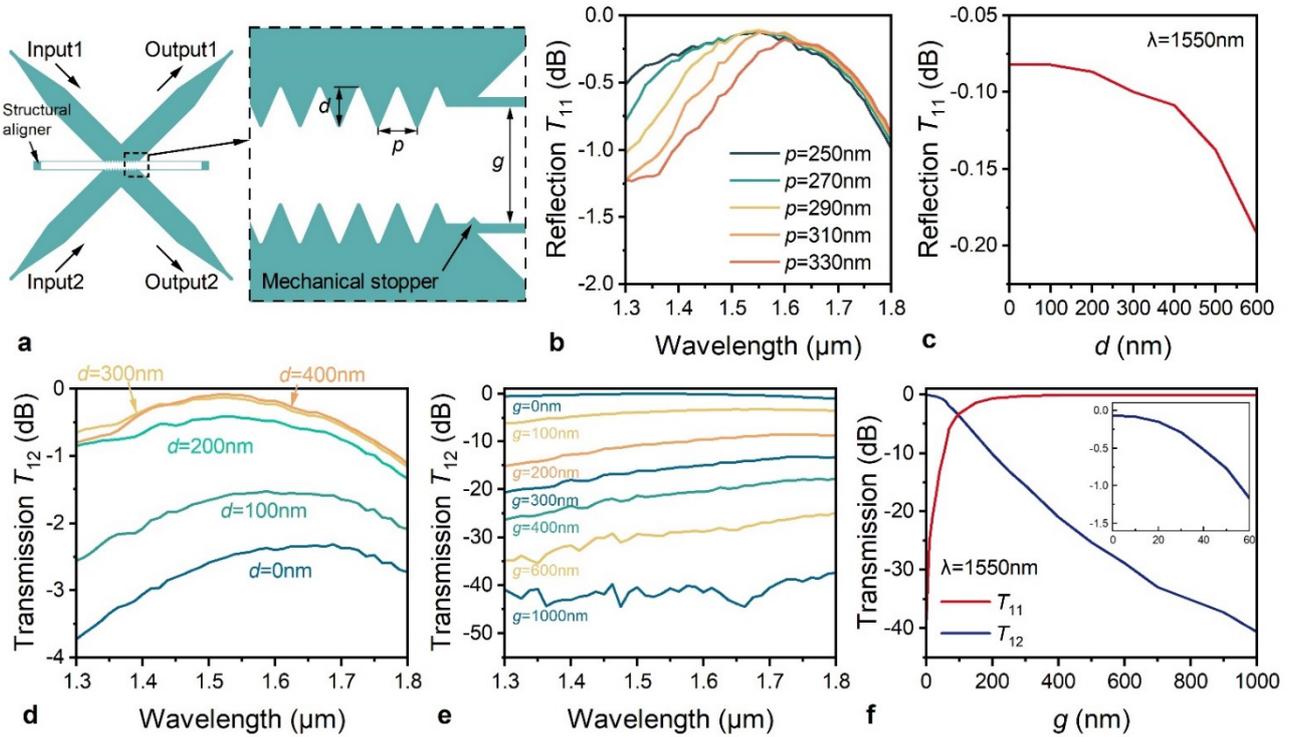

**Fig. S2 | Split waveguide crossing (SWX) design. a**, Schematic diagram of the SWT facets in the SWX. $p$ and $d$ represent period and depth of the SWT structures respectively. $g$ represents the gap width between the two facets. **b**, Reflection loss $T_{11}$ with different SWT period ($d$ = 300 nm). **c**, Reflection loss $T_{11}$ at 1550 nm with different SWT depth ($p$ = 290 nm). **d**, Transmission $T_{12}$ of the switch in ON state ($g$ = 20 nm) with different SWT depth ($p$ = 290 nm). **e**, Transmission $T_{12}$ with different gaps between the two SWT facets. **f**, Optical transmission $T_{11}$ and $T_{12}$ at 1550 nm with different gaps between the two facets. Inset shows the transmission $T_{12}$ for gaps $g$ narrower than 60 nm.

Note that the gap also plays an important role for the SWX. Figure S2e shows the transmission $T_{12}$ (or $T_{21}$) for the SWX with different gap widths, which proves that the designed SWX exhibits broadband operation. Figure S2f shows the transmissions $T_{11}$ (or $T_{22}$) and $T_{12}$ (or $T_{21}$) at the central wavelength 1550 nm with different gap width $g$. Apparently, the crosstalk can be minimized in the OFF state by choosing a wide gap of hundreds of nm and beyond, which however requires a large displacement and hence high actuation voltages. On the other hand, a small gap of tens of nanometers and below is required to lower the crosstalk in the ON state, in which case the fabrication of the mechanical stoppers should be precise to avoid stiction. Taking all the aforementioned trade-offs into account, we choose the gap widths in the OFF and ON states as $g = 900$ nm (designed initially) and $g = 15$ nm (determined by the mechanical stoppers), respectively.

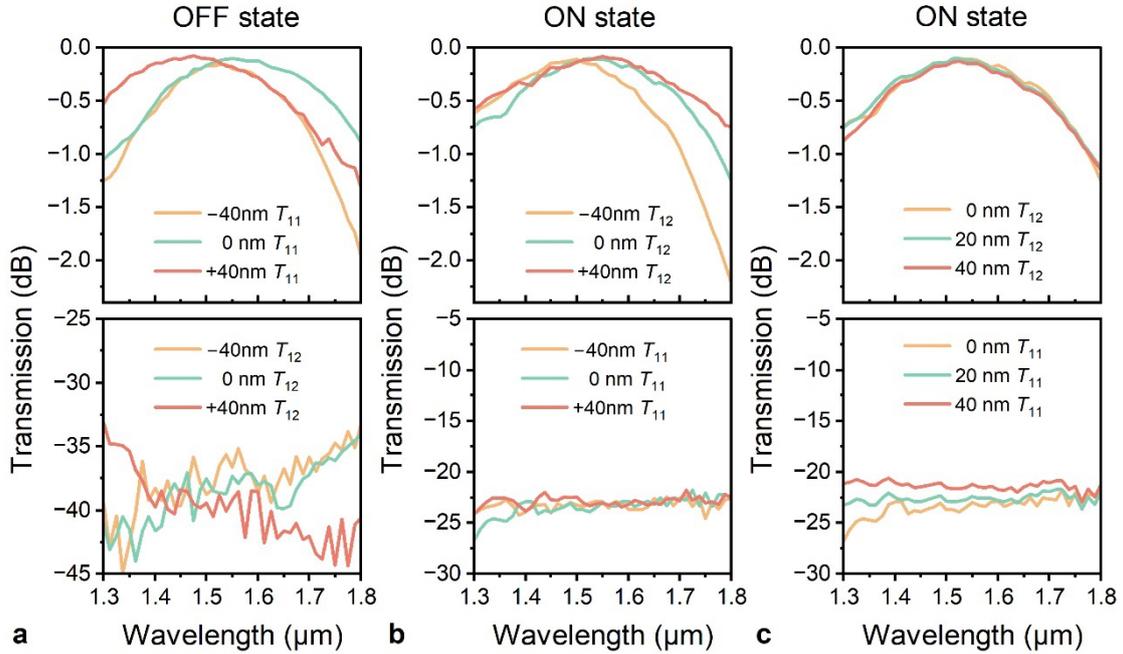

**Fig. S3 | Tolerance analysis of the SWX.** Performance changes of the **a,** OFF state, and **b,** ON state, caused by the deviation of the waveguide width. **c,** Performance changes of the ON state caused by $z$ misalignment of the SWT facets.

Ideally, in OFF and ON state, the designed SWX performs 0.1–0.52 dB excess loss with < −37.1 dB crosstalk and 0.1–0.47 dB excess loss with < −22.5 dB crosstalk respectively in the operating bandwidth of 1400–1700 nm. Here, Tolerance of the SWX is analyzed in the operating bandwidth of 1400–1700 nm. Figs. S3a and S3b show the calculated transmissions $T_{11}$ and $T_{12}$ for the SWX with different width variations. In the OFF state, when the width variation is +/− 40nm, the highest excess loss in the operating bandwidth increases to 0.78 dB at maximum and the crosstalk $T_{12}$ increases to < −35.2 dB. In the ON state, with 40nm wider or narrower waveguides, the highest excess loss in the operating bandwidth increases to 0.94 dB at maximum and the crosstalk $T_{11}$ increases slightly to −22 dB. In addition, the excess loss $T_{12}$ in ON state caused by the misalignment in the $z$ direction was estimated in theory. As shown in Fig. S3c, the excess loss $T_{12}$ in the ON state increases slightly to 0.13 − 0.52 dB with a 40nm $z$-misalignment. Evidently, the crosstalk increases with $z$-misalignment as well, especially at the short wavelength. Fortunately, when the $z$-misalignment is 40 nm, the crosstalk increases to < −21 dB slightly, which is acceptable. Owing to the meticulous mechanical designs (as described in Section 3), both $z$-misalignment and $x$-misalignment should be very small. Particularly, in our experiments no crosstalk larger than −20 dB was observed, which verifies the robustness of the present structural design.

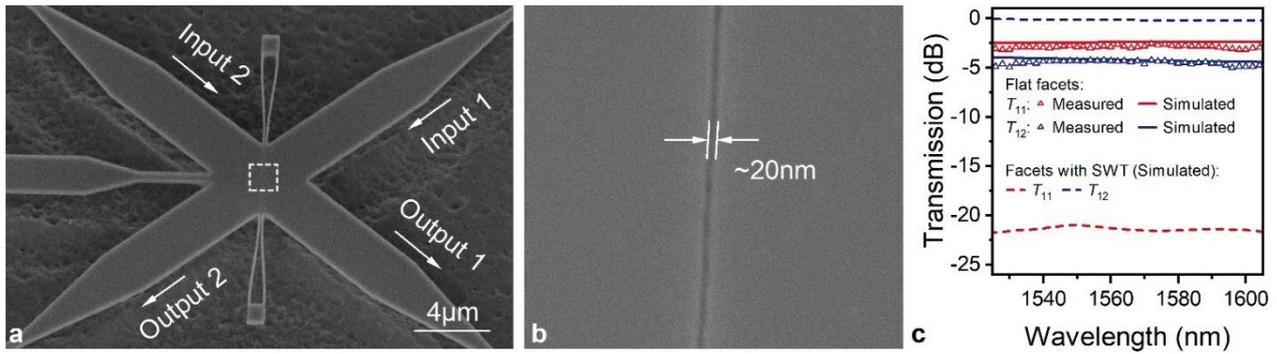

**Fig. S4 | Simulated and experimental results of a SWX switch with flat reflecting facets. a**, Scanning electron microscope (SEM) image of the SWX in stiction. **b**, Close-up view of the wedge-shaped gap between the two stictioned reflectors (areas within the white dashed box in **a**). **c**, Simulated and experimental transmission spectra of the stictioned SWX.

To further justify the importance of the SWT structures, a SWX switch with flat reflecting facets (without SWT structures) was also fabricated and characterized as a testing device, as shown in Fig.S4. Due to the lack of mechanical stoppers, when the two facets engaged, they were stictioned tightly by the van der Waals force, as shown by the SEM picture in Fig. S4a. Figure S4b is a close-up view of the area around the gap in Fig. S4a. Since the sidewalls are not vertically perfectly (The testing device was fabricated on a different wafer and the sidewall inclination is larger than our formal device), there is a wedge-shaped gap formed with a small gap width of ~20 nm (full width at the half height). The measured transmissions $T_{11}$ and $T_{12}$ of this stictioned switch are shown as triangular symbols in Fig. S4c. Due to the existence of the small gap between the two facets, there is still ~3 dB reflection for the stictioned SWX. The solid lines in Fig. S4c represents the simulated transmissions $T_{11}$ and $T_{12}$ of an SWX without SWT structures when the gap width is 20 nm, which agrees well with the measurement. Therefore, mechanical stoppers should be introduced to avoid any stictions.

Accordingly, we further introduce SWT structures at the reflecting facets, which is necessary to improve the transmissions in the ON state. Fig. S4c also shows the calculated transmissions $T_{11}$ and $T_{12}$ of an SWX switch with a 20-nm-wide gap between the SWT facets (see the dashed lines). It can be seen that the excess loss is reduced to 0.15 dB and the extinction ratio is improved to 21dB at 1550nm. Obviously, the SWT structures greatly help enhance the switch performance even with the air gap.

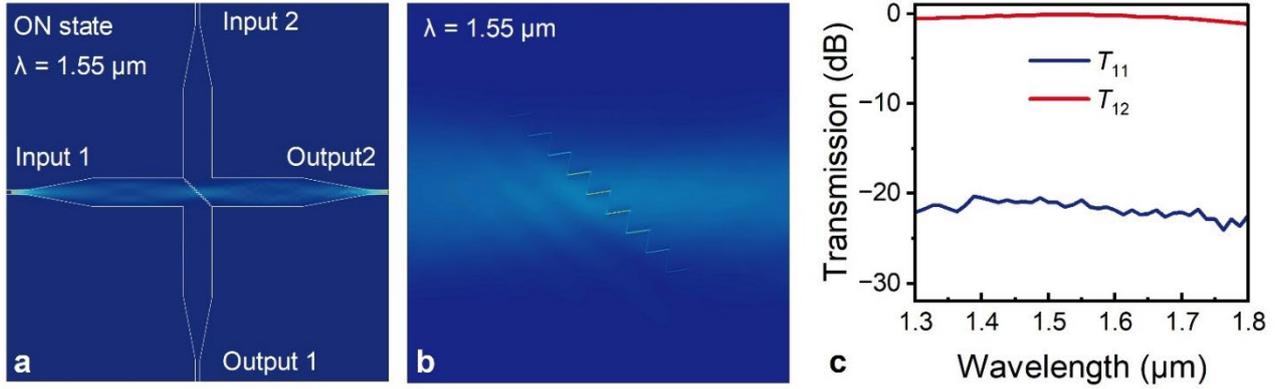

**Fig. S5. Simulation of the SWT with _p_ = 290nm, _d_ = 200nm and _g_ = 10nm in cross state. a**, Light field propagation in the SWX in ON state at 1550 nm. **b**, Close-up view of the light propagation at the SWT facets. **c**, Transmission spectra of the switch in ON state.

It can be seen in Section 2 that the switch performance in ON state can be improved (especially the crosstalk) by decreasing the air gap. However, due to the sidewall inclination, it is challenging to further decrease the air gap in the engaged SWX, as shown in Fig. S4. When nearly vertical sidewalls can be fabricated with improved processes, the gap can be further minimized to sub-10 nm. In this case, it is allowed to use the SWT with a smaller aspect ratio (_d_ / _p_), which can be fabricated easier. Fig. S5 shows the simulation results for the SWT with _p_ = 290 nm and _d_ = 200 nm for the case of _g_ = 10 nm. It can be seen that the excess loss is as low as 0.1–0.5 dB and the crosstalk is < –21 dB over the 1400 – 1700 nm wavelength range. More importantly, such a structure can be readily fabricated with commercial silicon photonics foundry process with a feature size of 180/130 nm. As a summary, it is desired to have vertical sidewalls for better performances and easier fabrication.

**Section 3: Mechanical Design**

In our device design, we prioritized a low actuation voltage. Thus, the mechanical structures should be sufficiently flexible. All the mechanical structures were designed with finite element method (FEM) simulation.

The entire mechanical structure consists of three parts, i.e., the movable waveguide, the structural aligner and folded springs, and the shuttle beam connected with the movable waveguides and the springs.

The movable waveguides include a pair of meandering spring waveguides, which connect the suspended movable reflector and the mode converters, are the most rigid part in the mechanical structures. They contribute dominantly to the overall spring constant of the entire mechanical structure. Therefore, they are the key to achieve a low actuation voltage.

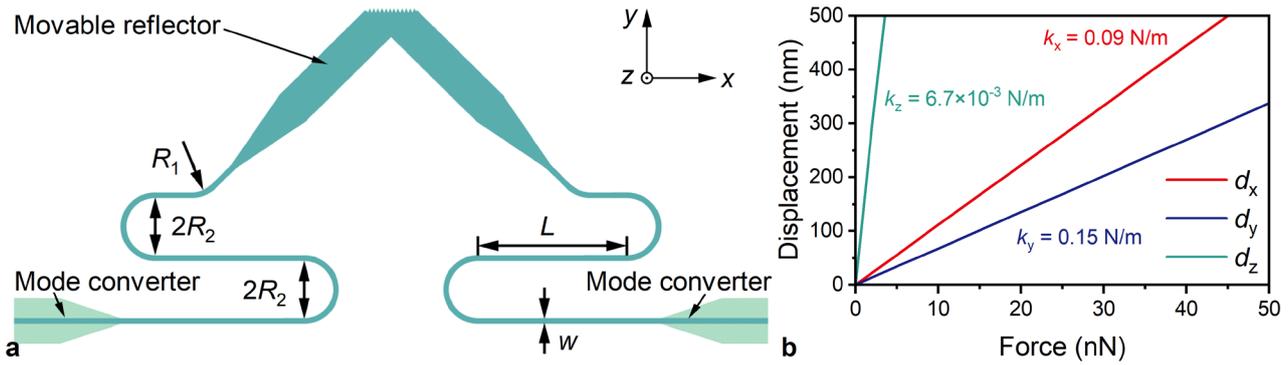

**Fig. S6 | Simulation of the meandering spring waveguides. a**, Schematic diagram of the movable waveguides which include the movable reflector and meandering spring waveguides. **b**, Simulated spring constants of the movable parts in $x$, $y$ and $z$ directions.

Figure S6a shows the schematic diagram of the movable waveguides which include the movable reflector and spring waveguides. Two 180° waveguide bends are incorporated within each spring waveguide to be more flexible. We choose the core width as $w = 400$ nm to support the $TE_0$ mode well and afford excellent flexibility. Specifically, we choose the bending radii as $R_1 = 5$ μm and $R_2 = 3$ μm for negligible bending loss and $L = 15$ μm was adopted for excellent flexibility within a compact footprint.

Figure. S6b shows the calculated displacement of the movable reflector ($d_x$, $d_y$ and $d_z$) with different actuation force in the $x$, $y$ and $z$ directions. The spring constants $k_x$, $k_y$ and $k_z$ in the $x$, $y$, and $z$ directions can be derived from the slopes of the displacement–force curves, and one has $k_x = 0.090$ N/m, $k_y = 0.15$ N/m, and $k_z = 6.7 \times 10^{-3}$ N/m. Here the spring constant $k_z$ is quite low because the load is applied on the reflection facet in the simulation, which induces a large rotational force moment with respect to the fixed ends of the mode converters. To suppress the undesired displacement in the $z$ direction, additional aligners (elaborated below) were employed to stabilize the movable reflector.

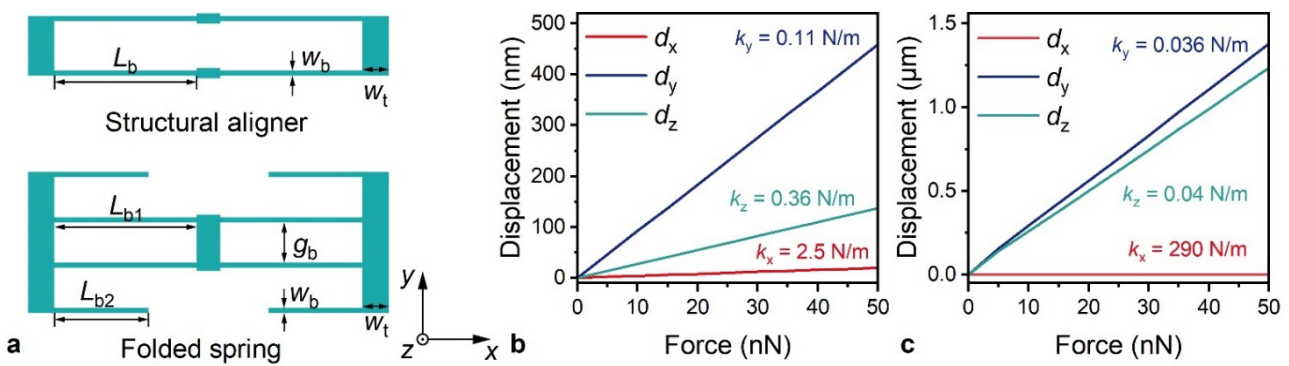

**Fig. S7 | Simulation of the springs. a**, Schematic diagram of the structural aligner and folded spring. **b**, Simulated spring constants of the **b** structural aligner and **c** folded spring in different directions.

The second part of the mechanical structures is the structural aligner and folded springs.

Figure S7a shows the schematic diagram of the structural aligner, which are used to minimize any misalignment between the two SWT facets, especially in the $z$ direction. Meanwhile, they should not significantly increase the rigidity of the movable reflector in the $y$ direction, in order to maintain a low actuation voltage. Therefore, the spring constants of the structural aligners were engineered to be high in the $x$ and $z$ directions but low in the $y$ direction. Based on the considerations above, we chose the following structural parameters: $L_b$ = 7 μm, $w_b$ = 0.1 μm, and $w_t$ =1 μm. Fig. S7b shows the simulated displacement–force curves of the structural aligner in different directions. The spring constants in different directions are derived as: $k_x$ = 2.5 N/m, $k_y$ = 0.11 N/m, $k_z$ = 0.36 N/m, exhibiting high stability in the $x$/$z$ directions and flexibility in $y$ direction.

For the folded spring (see Fig. S7a), which are used to support the entire mechanical structures and ensure that the movable parts move only along the shuttle beam in the $y$ direction, its spring constants were desired to be high in the $x$ direction and low in the $y$ direction. We chose the beam lengths $L_{b1}$ = 30 μm and $L_{b2}$ = 22 μm, the beam width $w_b$ = 0.2 μm, and the truss width $w_t$ = 1 μm. To avoid any contact between the adjacent beams, the gap width $g_b$ = 1.5 μm was adopted to allow 0.9-μm displacement in the $y$ direction. The simulated displacement–force curves of the folded spring with these design parameters are shown in Fig. S7c. The spring constants in different directions are then derived as: $k_x$ = 290 N/m, $k_y$ = 0.036 N/m, $k_z$ = 0.04 N/m.

For the shuttle beam connected with the movable waveguides and the springs, the width was designed to be 6 μm to be robust against buckling. The shuttle beam is perforated to achieve better rigidity with less mass, hence less inertia and faster response. Meanwhile, the perforation facilitates the undercut process during the hydrofluoric (HF) acid vapor etching. The movable waveguides and the springs are effectively connected in parallel in the $y$ direction, thus the total spring constant is derived as $k_{y\text{-total}}$ = $k_{y1}$ + $k_{y2}$ + $2k_{y3}$ = 0.15 + 0.11 + 2 × 0.036 = 0.33 N/m.

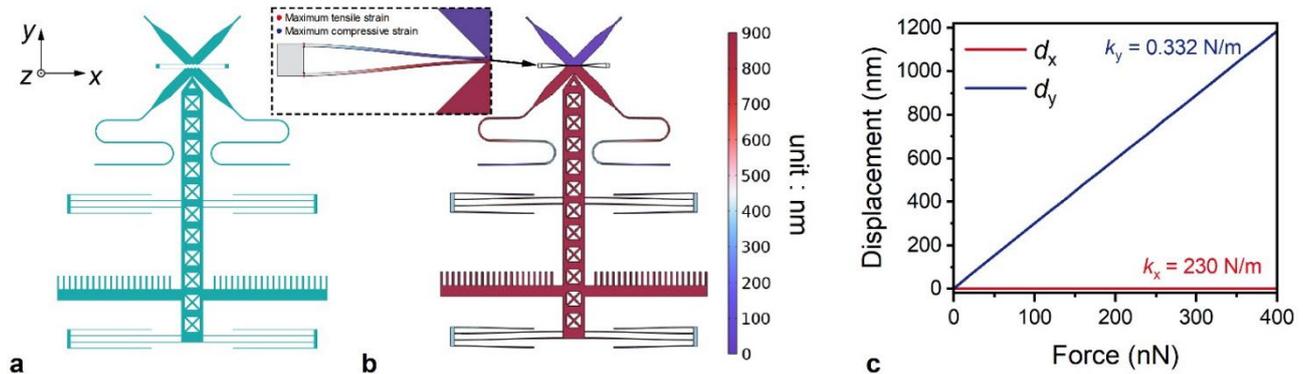

**Fig. S8 | Simulation of the entire mechanical structures. a,** Schematic diagram of the mechanical structures. **b,** $y$-direction displacement distributions of the entire structures. The inset shows the position where the maximum strains locate. **c,** Spring constant in $x$ and $y$ directions ($k_x$ and $k_y$) of the mechanical structures.

To verify the design methodology, the entire mechanical structures shown in Fig. S8a were simulated and Fig. S8b shows the $y$-direction displacement distributions of the mechanical structures in the ON state. The maximum tensile and compressive strains are ~±3×10⁻³ locate at the structural aligners and the detailed positions are shown in the inset of Fig. S8b. Figure S8c shows the displacement in the $x$ and $y$ direction ($d_x$ and $d_y$) with different actuation forces. The spring constant in the *y* direction $k_y$ is calculated as 0.332 N/m, which agrees well with the derived one (i.e., 0.33 N/m). Besides, the spring constant in the *x* direction $k_x$= 230 N/m, indicating the high stability in the *x* direction.

## Section 4: Electrostatic Comb Design

The electrostatic combs were designed to realize low actuation voltage and high lateral stability. As shown in Fig. S9, a single unit cell of the electrostatic combs was considered. Theoretically, the electrostatic force $F_x$ in the *x* direction is zero because the gaps between the adjacent comb fingers are uniform (i.e., $d_{c1} = d_{c2}$). However, either the fabrication error or any external perturbation will make $d_{c1} \neq d_{c2}$. As a result, considerable $F_x$ arises and creates instabilities in the *x* direction of the structures. For our device, the stability in the *x* direction is critical because the SWT structures engage with a nm-scale gap in the ON state. Otherwise, any lateral contact may cause stictions.

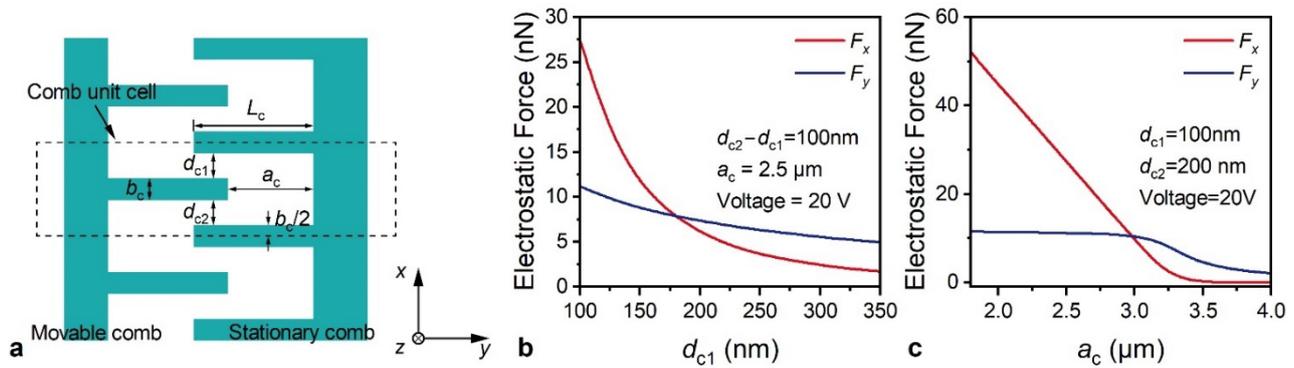

**Fig. S9 | Simulation of the electrostatic combs. a**, Schematic diagram of the electrostatic combs, a single unit cell is considered in the simulation. **b**, Electrostatic force in *x* and *y* directions ($F_x$, $F_y$) as functions of $d_{c1}$, assuming $d_{c2} - d_{c1}$= 100 nm and voltage = 20 V. **c**, $F_x$ and $F_y$ as functions of $a_c$, assuming $d_{c1}$ = 100 nm, $d_{c2}$ = 200 nm and voltage = 20 V.

For the present design, we choose the comb-finger length $L_c$ and the comb-finger width $b_c$ as $L_c$ = 3.2 μm and $b_c$ = 0.5 μm, respectively. To achieve high stability in the *x* direction, the gap widths ($d_{c1}$, $d_{c2}$) and the distance $a_c$ should be designed to minimize the force $F_x$. Figure S9b shows the relationship between the electrostatic force motivated by a single pair of combs and the gap width ($d_{c1}$, $d_{c2}$), by assuming $d_{c2}-d_{c1}$ = 100 nm and $V_{bias}$= 20 V. With the gap width decreases, the force $F_x$ increases quickly while the force $F_y$ increases much more slowly. Figure S9c shows the electrostatic force motivated by a single pair of combs as the distance $a_c$ varies, when assuming $d_{c1}$ = 100 nm, $d_{c2}$ = 200 nm and $V_{bias}$ = 20 V. If the fingers of the two combs still overlap in the *y* direction, corresponding to the case of $a_c$ < 3.2 μm, the force $F_x$ decreases fast and the force $F_y$ almost keeps constant with the increase of $a_c$. If the fingers of the two combs separate in the *y* direction (i.e., $a_c$ > 3.2 μm), the force $F_x$ decreases to zero and the force $F_y$ decreases significantly with the increase of $a_c$. Finally, we choose $d_{c1}$ = $d_{c2}$ = 300nm and $a_c$ = 3 μm to obtain strong $F_y$ as well as weak $F_x$. To lower the actuation voltage, the number of comb fingers can be increased. However, constrained by the flexural deformation of the movable comb handle and the overall size of the device, we choose 40 unit-cells of the electrostatic combs in total.

With this design, each pair of comb fingers contributes a force of 7.4 nN to switch the device to the ON state, when a pull-in voltage of 22V is applied.

## Section 5: Resonance Frequency of the Mechanical Structures

The fundamental resonance frequency of the mechanical structures is estimated as a simple harmonic oscillator (SHO) here. The total volume and mass of the movable mechanical structures are $V$ = 192.42 $\mu m^3$ and $m$ = 4.5 × $10^{-13}$ kg, respectively (with the density of $\rho$ = 2.33 × $10^3$ kg/$m^3$). With the spring constant of $k_y$ = 0.332 N/m estimated above, the mechanical structure has a resonance frequency of $f$ = 0.14 MHz, derived from $\omega = 2\pi f = \sqrt{k/m}$. It agrees well with the calculated eigenfrequency of 0.152 MHz from the simulation of the entire mechanical structures. Such a resonance frequency indicates that the highest switching frequency is up to ~100 kHz, enabling μs-scale switching speed.

Additionally, the speed of our device is influenced by the contact resistance between the chromium and silicon, the resistance of silicon, the initial gap of the SWX, the spring constant of the mechanical structure and the applied drive voltage.

(a) The contact resistance between the chromium and silicon, as well as the resistance of silicon top layer, affects the effective charging circuit of the electrostatic comb. During device fabrication, we directly evaporate chromium on the silicon surface, leading to a Schottky contact and considerable contact resistance, compromising the temporal response of the switch. If Ohmic contact with lower resistance can be formed by heavily doping or annealing, faster switching speed can be obtained. However, we currently lack the experimental capability to carry out these specific enhancements.

(b) The initial gap determines the movement distance of the mechanical structure, a smaller initial gap results in a shorter movement distance, leading to faster temporal response, however, at the price of higher crosstalk in the OFF state.

(c) The spring constant of the mechanical structure is a critical factor influencing the temporal response. Increasing the elastic coefficient will enhance the mechanical resonance frequency of the device, thereby improving the temporal response, however, at the price of higher threshold driving voltages (Section V of the Supplement).

(d) The applied driving voltage is proportional to the square root of the thrust exerted on the mechanical structure, and a higher driving voltage leads to an increase in ON switching speed.

## Section 6: Stability of the Electrostatic Actuation

To analyze the lateral ($x$ direction) stability of the device, the total potential energy is calculated for the entire movable mechanical structures as function of the displacement in $x$ direction. Figure S10a shows the mechanical restoring force and potential energy of the mechanical structures, with linear and quadratic dependence on the *x*-displacement, respectively, which is typical for an SHO. A deep potential well can be observed, suggesting excellent stability in the *x* direction. Figure S10b shows the electrostatic force and the electric potential energy of the designed electrostatic combs for a sample of 15-V drive, where linear and quadratic dependence (approximately) of the *x*-displacement from −100 to 100 nm is observed, respectively. Without a potential well, the electrostatic combs alone are not stable in the *x* direction due to the pull-in effect. The electrostatic force and the electric potential energy are both proportional to the drive voltage squared. The total potential energy includes the mechanical potential and the electric potential, and Fig. S10c shows the total potential curves with different voltages. The potential well becomes shallower with increasing voltage, but a sufficiently deep potential well still exists with 200-V voltage, indicating the excellent stability in the *x* direction of our device.

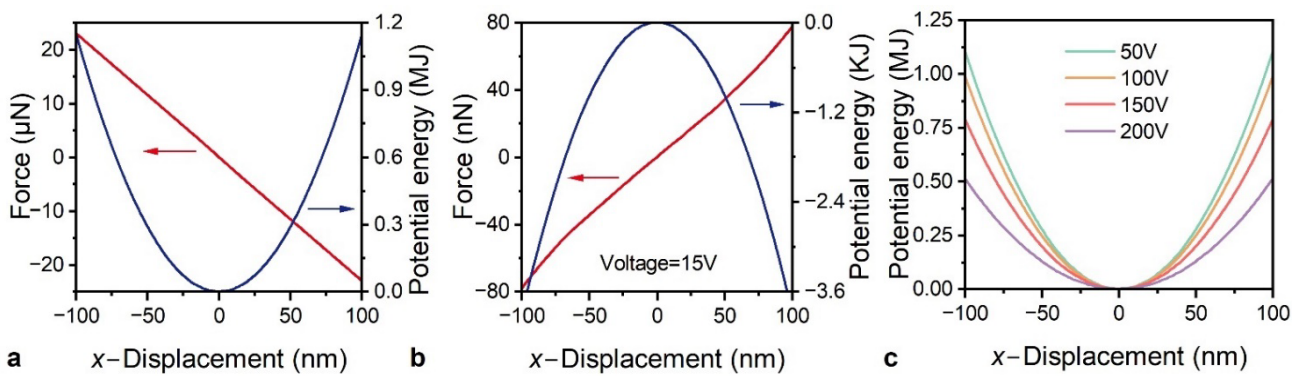

**Fig. S10 | Simulated potential energy curves. a**, Mechanical potential energy of our device in the *x* direction. **b**, Electric potential energy of the designed electrostatic combs for 15-V drive. **c**, Total potential energy with different drive voltages of our device.

### Section 7: Mechanical Shock Resistance

To investigate the resistance of the device against mechanical shocks, steady accelerations up to 100g (g is the gravitational acceleration) in different directions are applied to the entire mechanical structures in the OFF state for the simulation and the deformations in *x*, *y*, *z* directions $d_x$, $d_y$, $d_z$ are respectively calculated, as shown in Fig. S11b. Displacement below 3.2 nm is observed in all the simulated directions, which is far from damaging the device. The mechanical resonance ring-down process due to strong mechanical shocks can be well suppressed by gas damping. Therefore, with proper MEMS hermetic packaging[15] (e.g., with $N_2$ gas), the SWX switch will features superior resistance against mechanical shock.

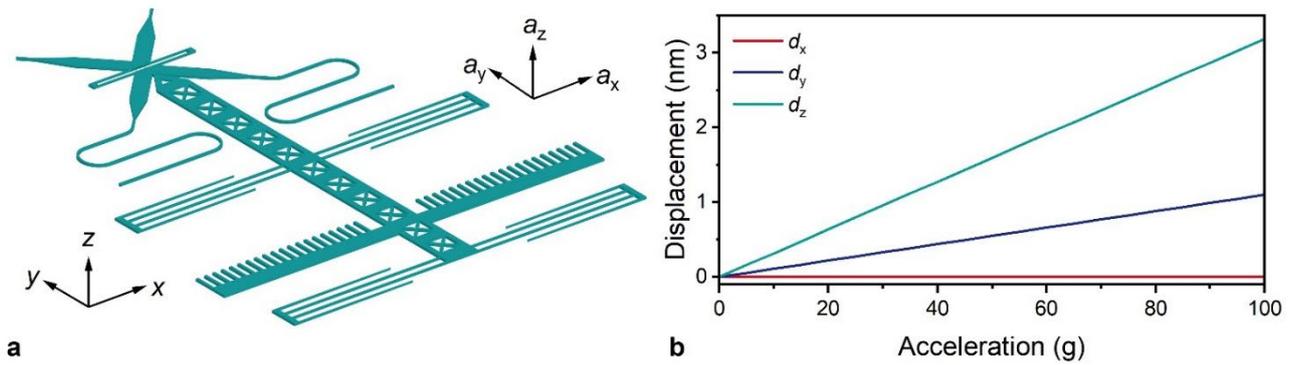

**Fig. S11 | Simulations of the mechanical shock resistance. a**, Schematic diagram of the simulation model. **b**, Displacement as a function of steady acceleration in different directions.

## Section 8: Theoretical Energy/Power Consumption of the Switch

To theoretically estimate the power consumption of the present switch, we calculate the mechanical and electric potential energy of the device in the ON state. When the switch is turned on, a displacement of 900 nm occurs. With a spring constant of $k_y = 0.33$ N/m, the mechanical potential energy can be derived to be 134 fJ. For the electric potential energy, the simulated capacitance of the electrostatic combs is ~2.04 fF. With a drive voltage of 20 V, the electrical potential energy can be derived to be 408 fJ. Therefore, the theoretical total switching energy is then 542 fJ, equal to the sum of the mechanical and electrical potential energy. Assuming the maximum switching frequency of 100 kHz, the maximum power consumption is calculated to be 54.2 nW.

In addition, the leakage current of the device which causes additional power consumption is measured, as shown in Fig. S12. We employed a sourcemeter (Keithley 2400) with resolution of 10 pA to perform the measurement. The I-V curve measured in the open-circuit state, shown in Fig. S12a, represents the noise current. While, the other one indicates the leakage current of our device. Obviously, the leakage current is below the noise floor and can be considered lower than 0.2 nA, resulting in an additional power consumption less than 4 nW, which is negligible compared to the theoretical maximum power consumption of 54.2 nW.

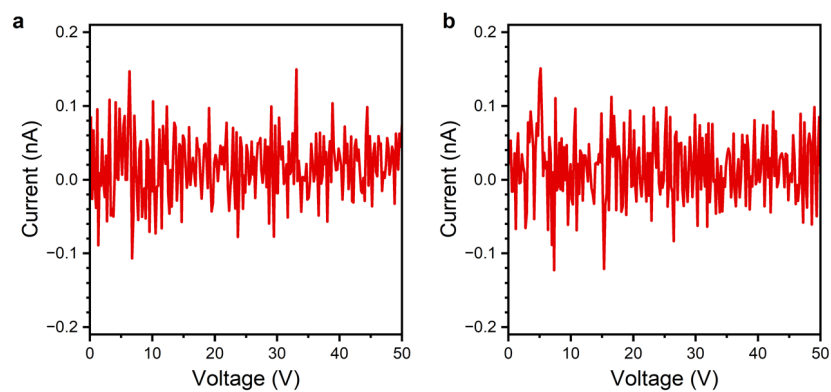

**Fig. S12 | Measurement of the leakage current. a,** results of open circuit state. **b,** results of the device.

**Section 9: Switch Operation Measurements**

Optical transmissions with different voltages were simulated and measured. The red solid line in Fig. S13a shows the simulated optical transmissions $T_{12}$ at 1550 nm. Evidently, as the drive voltage increases from 0 to 22 V (the threshold voltage), the transmission increases from 37.1 dB to 0.12 dB. After the threshold voltage is reached, the SWT facets engage and the transmission becomes constant at 0.12 dB, corresponding to the transmission $T_{12}$ in the ON state. The blue triangular symbols in Fig. S13a shows the measured data, which agrees with the simulation. The measured threshold voltage is 20V, which is slightly lower than the simulated one because the fabricated mechanical structures might be more flexible due to fabrication errors.

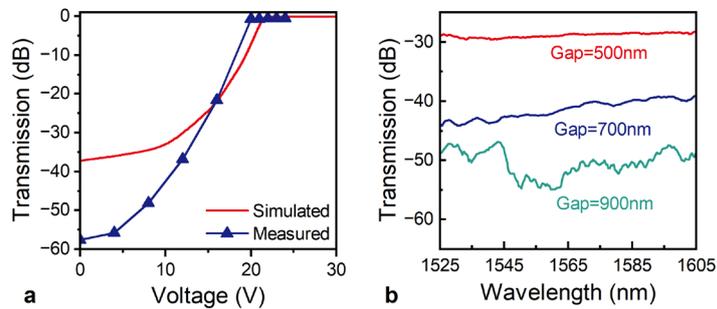

**Fig. S13 | Switch operation measurements with different voltages. a**, simulated and measured optical transmission $T_{12}$ at 1550 nm of the output port 2 with different drive voltage. **b**, measured optical transmission $T_{12}$ of switches with different initial gaps in the OFF state.

To investigate the effects of the initial gap between the SWT facets on the switch performance, switches with different initial gaps were fabricated and measured. Figure S13b gives the measured crosstalk of the switches in the OFF state, which decreases from < –29.3 dB to < –54.5 dB at the wavelength of 1550 nm when the initial gap increases from 500 nm to 900 nm.

**Section 10: Durability Test**

To investigate the durability of our switches, we have operated the switch over 1 billion cycles with a square-wave voltage at 20 kHz in our ~14-hour testing experiments. Because there is some slow drift of the alignment between the input/output fibers and the grating couplers, the transmission spectra were measured with once realignment every two hours. Figure S14 shows the measured seven groups of transmissions $T_{12}$ recorded. It can be seen that the results are repeated very perfectly (with a very small variation of ~0.1dB).

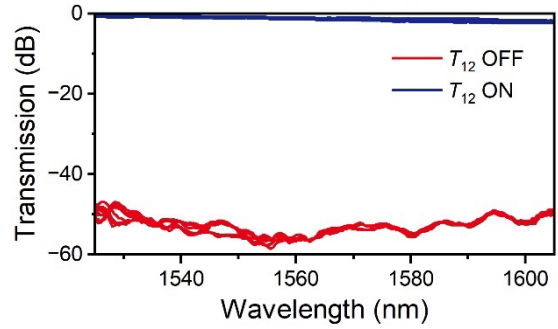

**Fig. S14 | Measurement results of the durability.** Seven groups of transmission spectra have been recorded over ~14 hours, with one group every 2 hours.

## Section 11 : Excess Loss Measurements of the Switch Components

To measure the excess loss of long routing waveguides, the mode-converter pairs, the circular 90°-bends, the Euler S-bends and the varied-width MMI waveguide crossings, a series of testing structures were designed and fabricated, as shown in Fig. S15a, b.

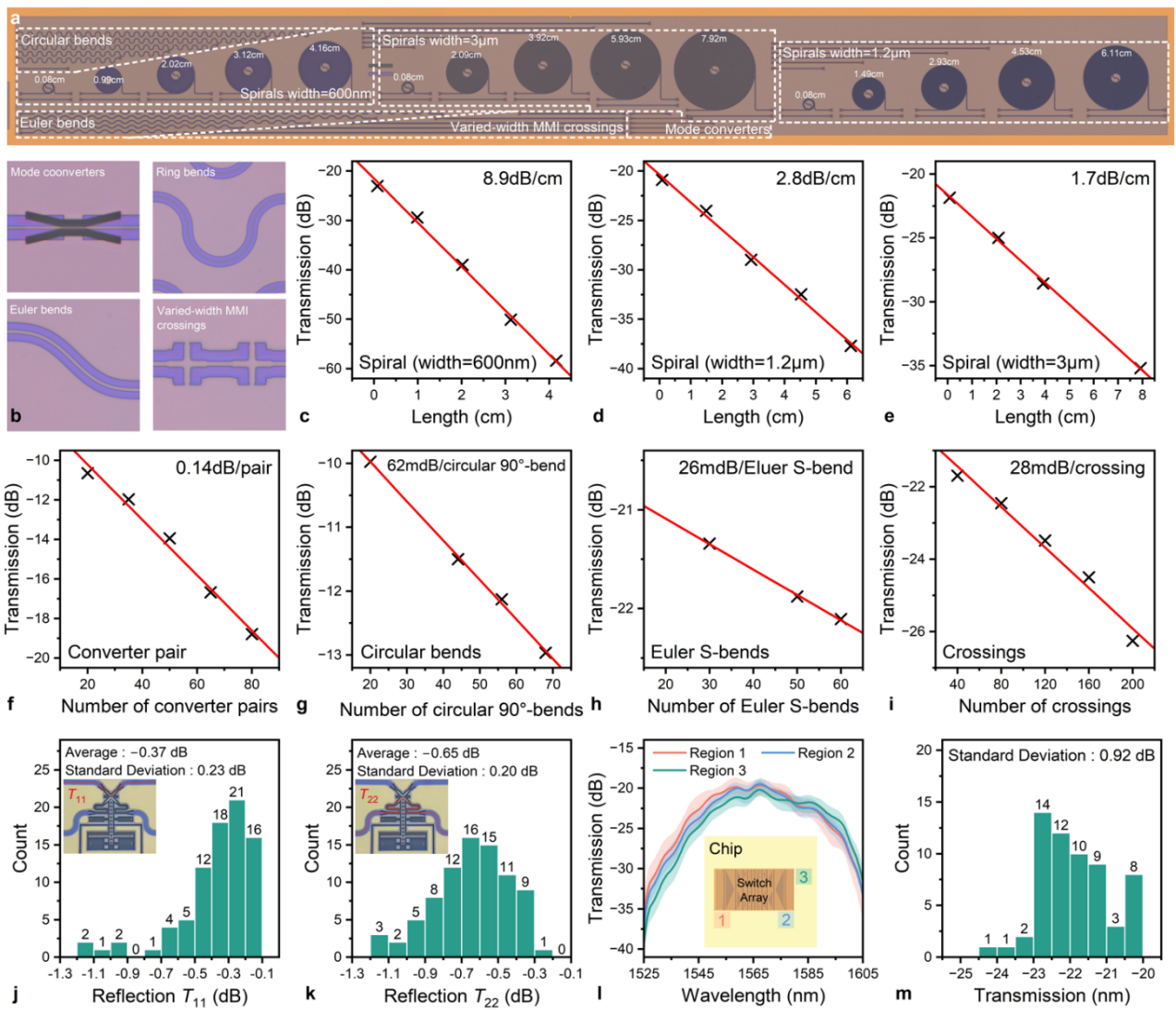

**Fig. S15 | Measurements of the cascaded passive optical test structures. a,** Microscope image of the optical test

structures. **b,** Close-up views of the mode-converter pair, circular 90°-bend, Euler S-bend and varied-width MMI waveguide crossing, respectively. **c–i,** Propagation/excess loss versus length of the routing waveguides, number of the mode-converter pairs, circular 90°-bends, Euler S-bends and varied-width MMI waveguide crossings, respectively. **j, k,** The statistics of the excess loss of the elementary switches. **l, m,** The statistics of the transmission spectra of the calibration waveguide.

From the linear fitting of the measurement results in Fig. S15c–i, the measured propagation losses of the routing waveguides with width = 600 nm, 1.2 μm, 3 μm are 8.9 dB/cm, 2.8 dB/cm and 1.7 dB/cm respectively. The measured excess losses of the mode converter-pair, the circular 90°-bend, the Euler S-bend and the varied-width MMI waveguide crossing are 0.14 dB, 62 mdB, 26 mdB and 28 mdB, respectively.

To characterize the uniformity of the fabrication and analyze the excess loss of the switch array, we survey 82 2×2 elementary switches and 60 calibration waveguides with grating couplers near the switch array.

The statistics of the transmission $T_{11}$ and $T_{22}$ at 1550 nm in the OFF state is shown in Fig. S15j, k. $T_{11}$ ranges from 0.1 dB to 1.13 dB and $T_{22}$ ranges from 0.23 dB to 1.13 dB. The measured average excess losses are 0.37 dB and 0.65 dB for $T_{11}$ and $T_{22}$ respectively, while the standard deviations are 0.23 dB and 0.2 dB for $T_{11}$ and $T_{22}$ respectively. $T_{22}$ is 0.28 dB larger than $T_{11}$ due to the extra waveguide bends, as shown in the inset of Fig. S15j, k.

## Section 12: Benes Switch Array Design and Excess Loss Analysis

The statistics of the transmission spectra of the calibration waveguides are shown in S15l, m. As shown in Fig. S15l, 20 calibration waveguides were characterized in Regions 1, 2, and 3, respectively. As shown in Fig. S15m, the transmissions of the calibration waveguides at 1550 nm range from –20 dB to –24.5 dB with a standard deviation of 0.92 dB.

The schematic diagram of a 64 × 64 switch based on Benes topology is shown in Fig. S16a. Ridge waveguides (150 nm partial etch) are used to connect the elementary switches to avoid being undercut during the HF etching process. In total, there are 352 elementary switches and 1824 waveguide crossings included. Figure S16b shows the optical microscope image of the connection shuffle between the first and second switch stages in our device. There are many waveguide bends introduced to achieve a compact footprint. Figure S16c shows the close-up view of the waveguide crossing and bends. In the connection shuffle, waveguides broadened to 1.2 μm are used instead of single-mode waveguides to reduce the propagation loss due to sidewall roughness. Therefore, waveguide bends in Euler curve should be used to suppress the excitation of high-order modes. In the connection shuffle, varied-width MMI waveguide crossings are used to achieve low loss and low crosstalk.

As shown in Fig. S16d, the Euler 45°-bend is defined by width ($w$), maximum curvature radius ($R_{max}$) and minimum curvature radius ($R_{min}$). We choose $w$ = 1.2 μm, $R_{max}$ = +∞, $R_{min}$ = 20 μm to achieve low excess loss, low excitation of high-order modes and compact footprint. To improve computational accuracy at the perfectly-matched-layer (PML) boundary, an S-bend consisting of two Euler 45°-bends is simulated, as shown in Fig. S16e. The excess loss of the S-bend is 3 mdB at 1550 nm and the high-order mode excitation is < –30 dB in the wavelength range of 1500 – 1600 nm. The measured excess loss of the Euler S-bend (Section 11 of the Supplement) is higher than the simulated one due to the propagation loss.

As shown in Fig. S16f, the varied-width MMI waveguide crossing consists of four arms that are four-fold rotationally symmetric about the center point. The arm of the varied-width MMI waveguide crossing is designed to be axisymmetric and it can be defined by the length ($L$) and the widths ($w_1$, …, $w_{13}$) at the 13 equidistant positions along the length, as shown in Fig. S16g. Based on the 13 widths, the varied-width edges are smoothly interpolated with cubic splines. For such a multiparametric optimization problem, we employ the particle swarm optimization (PSO) algorithm. After more than 120 iterations with FDTD method, optimal parameters of $L$ = 5.74 µm, ($w_1$, …, $w_{13}$) = (1.200, 1.521, 1.772, 1.792, 1.704, 1.711, 1.642, 1.779, 1.744,1.871, 2.078, 2.075, 0.700) µm are obtained, and the simulated results of the crossing are shown in Fig. S16h. The excess loss of the crossing is 4 mdB at 1550 nm and the crosstalk is < –38 dB in the wavelength range of 1500 – 1600 nm. Inset of Fig. S16h shows the simulated light propagation in the designed crossing at 1550 nm. The measured excess loss of the crossing (Section 11 of the Supplement) is higher than the simulated one due to fabrication errors and propagation loss. The simulated ultralow loss is experimentally validated in our previous work[16].

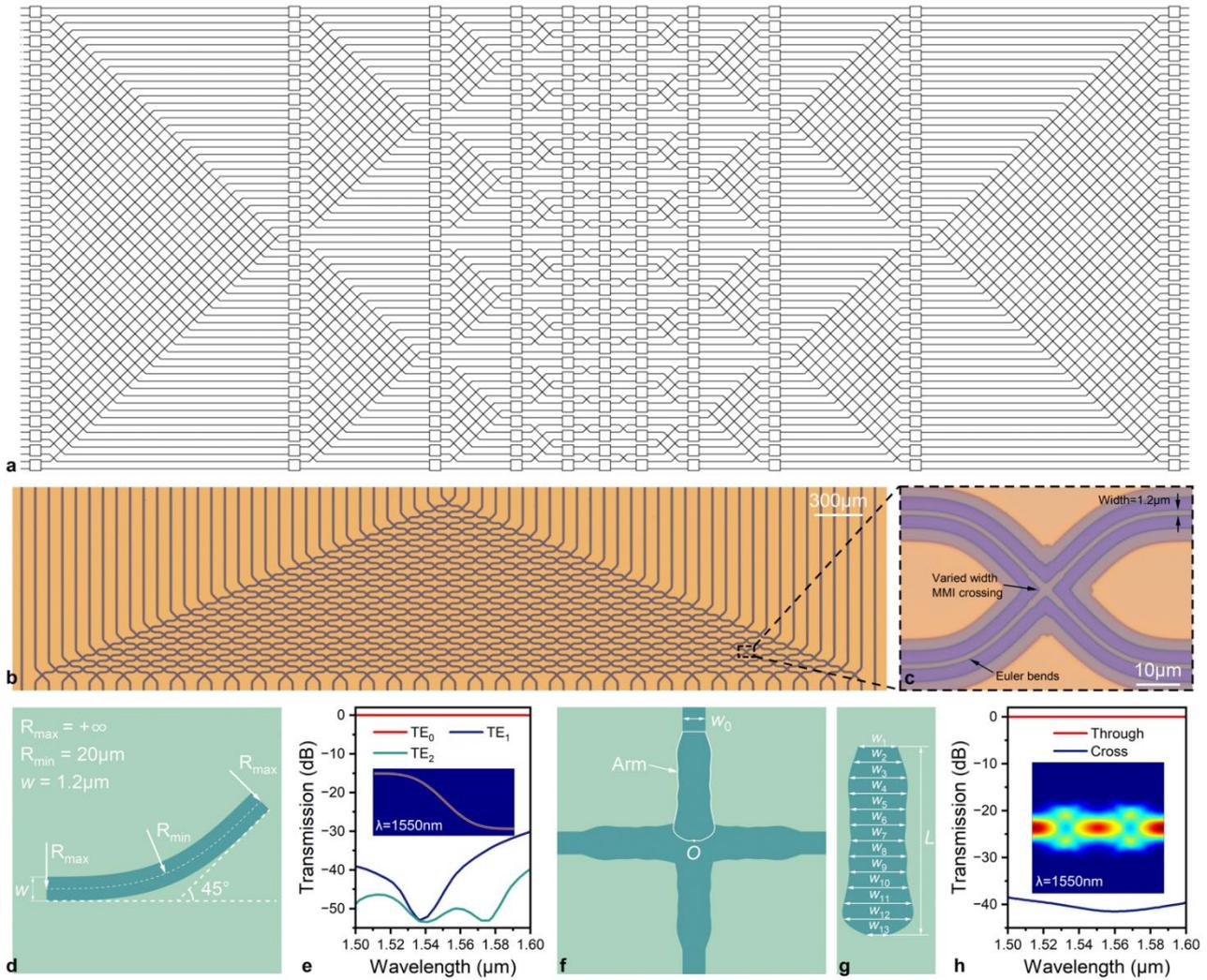

**Fig. S16 | Benes switch array design. a**, Schematic diagram of the 64 × 64 switch array based on Benes topology. **b**, Optical microscope image of the connection shuffle between the first and second switch stages. **c**, Close-up view of the waveguide crossing and Euler bends in the connection shuffle. **d**, Schematic diagram of the Euler 45°-bend. **e**, Simulated excess loss and high-order mode excitation of the Euler S-bend consisting of two centrosymmetrical Euler 45°-bends. Inset shows the light propagation at 1550 nm. **f**, Schematic diagram of the varied-width MMI waveguide crossing. **g**, Close-up view of the arm in the varied-width MMI waveguide crossing **h**, Simulated excess loss and crosstalk of the varied-width MMI waveguide crossing. Inset shows the light propagation at 1550 nm.

The fabrication tolerance for the varied-width MMI waveguide crossing is also analyzed in the wavelength range of 1500–1600 nm. Figure S17a, b shows the changes of the transmission spectra due to lithography errors. When the width deviation is ± 40 nm, the excess loss is still 3–31 mdB and the crosstalk is < −35 dB. Figure S17c, d shows the changes of the transmission spectra due to etching-depth errors. The excess loss is 3–16 mdB and the crosstalk is < −37 dB if the etching-depth varies with ± 10 nm.

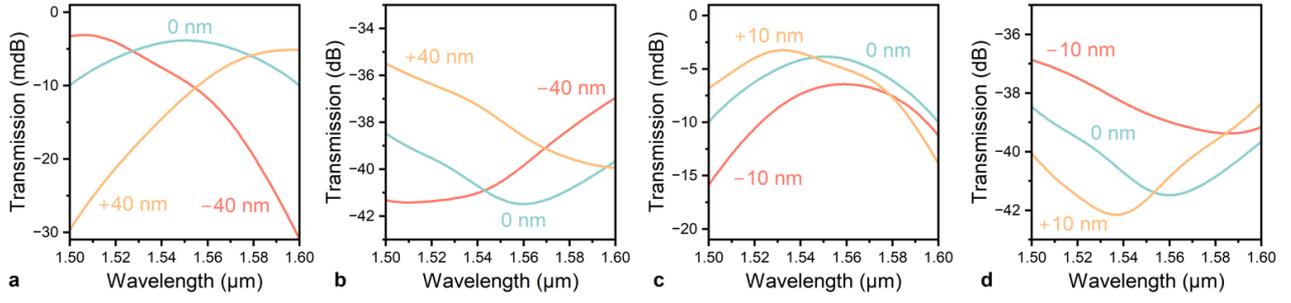

**Fig. S17 | Fabrication tolerance of the ridge-waveguide crossing. a, b,** Lithography tolerance. **c, d,** Etching-depth tolerance. A negative sign means the actual size becomes smaller, while a positive sign means the opposite.

In a Benes topology, input Port $i$ ($i$ = 1–64) is routed to the output Port $i$ in the all-OFF state. Therefore, transmission of the 64 optical paths correspond to $T_{i,j}$ ($i = j$). To analyze the excess loss of the fabricated switch array in all-OFF state, we survey the numbers of transmission $T_{11}$ for switch cells ($N_1$), transmission $T_{22}$ for switch cells ($N_2$), waveguide crossings ($N_c$), Euler S-bends ($N_b$) and the length of routing waveguides with width = 600 nm ($L_1$), 1.2 μm ($L_2$) and 3 μm ($L_3$) included in different optical paths and the results are shown in Table 2. All the SWX structures in the switch array are considered to have uniform excess loss of 0.37 dB ($T_{11}$) and 0.65 dB ($T_{22}$) at 1550 nm according to the statistics in Section 11. Meanwhile, the measured results with well fitted linearity in Section 11 also shows that the excess loss of varied-width MMI waveguide crossings and Euler S-bends is uniform and can be considered as 28 mdB/crossing and 26 mdB/bend in the switch array. The excess loss at 1550 nm of different optical paths is estimated with the measured results mentioned above, as shown in Table 2. The measured excess loss at 1550 nm is listed in the table as well, which agrees well with the estimated results for most of the paths.

Table 2. Excess loss analysis of the fabricated 64 × 64 Benes switch array (all-OFF)

| Path ID | $N_1$ | $N_2$ | $N_c$ | $N_b$ | $L_1$ (μm) | $L_2$ (μm) | $L_3$ (μm) | Estimated excess loss (dB) | Measured excess loss (dB) |
|---|---|---|---|---|---|---|---|---|---|
| 1 | 11 | 0 | 0 | 0 | 0 | 0 | 6428 | 5.16 | 10.63 |
| 2 | 9 | 2 | 62 | 62 | 0 | 320 | 3767 | 8.71 | 9.61 |
| 3 | 9 | 2 | 32 | 32 | 0 | 480 | 4494 | 7.26 | 10.43 |
| 4 | 7 | 4 | 90 | 90 | 0 | 710 | 2229 | 10.63 | 11.32 |
| **5** | **9** | **2** | **20** | **20** | **0** | **640** | **4590** | **6.67** | **42.03** |
| 6 | 7 | 4 | 74 | 74 | 0 | 940 | 2395 | 9.86 | 11.00 |
| 7 | 7 | 4 | 48 | 48 | 0 | 870 | 3377 | 8.60 | 9.08 |
| 8 | 5 | 6 | 98 | 98 | 0 | 870 | 1623 | 11.56 | 12.62 |
| 9 | 9 | 2 | 20 | 20 | 0 | 800 | 4264 | 6.66 | 8.05 |
| 10 | 7 | 4 | 66 | 66 | 0 | 800 | 2650 | 9.43 | 10.42 |
| 11 | 7 | 4 | 44 | 44 | 0 | 1100 | 3121 | 8.40 | 9.82 |
| 12 | 5 | 6 | 86 | 86 | 0 | 1100 | 1648 | 10.98 | 11.87 |
| 13 | 7 | 4 | 36 | 36 | 0 | 1030 | 3472 | 8.01 | 9.28 |
| 14 | 5 | 6 | 74 | 74 | 0 | 1030 | 2139 | 10.40 | 10.97 |
| 15 | 5 | 6 | 56 | 56 | 0 | 1030 | 2771 | 9.53 | 10.53 |

| 16 | 3 | 8 | 90 | 90 | 0 | 1030 | 1578 | 11.73 | 11.71 |
|---|---|---|---|---|---|---|---|---|---|
| 17 | 9 | 2 | 32 | 32 | 0 | 1260 | 3217 | 7.26 | 8.92 |
| 18 | 7 | 4 | 62 | 62 | 0 | 1260 | 2164 | 9.26 | 10.03 |
| 19 | 7 | 4 | 48 | 48 | 0 | 1260 | 2655 | 8.59 | 9.43 |
| 20 | 5 | 6 | 74 | 74 | 0 | 1260 | 1743 | 10.40 | 11.09 |
| **21** | **7** | **4** | **44** | **44** | **0** | **1561** | **2495** | **8.43** | **25.36** |
| **22** | **5** | **6** | **66** | **66** | **0** | **1561** | **1723** | **10.04** | **25.44** |
| 23 | 5 | 6 | 56 | 56 | 383 | 1561 | 3274 | 10.11 | 9.77 |
| 24 | 3 | 8 | 74 | 74 | 377 | 1561 | 2643 | 11.53 | 11.07 |
| 25 | 7 | 4 | 48 | 48 | 370 | 1190 | 3925 | 9.11 | 8.95 |
| 26 | 5 | 6 | 62 | 62 | 363 | 1190 | 3434 | 10.34 | 10.51 |
| 27 | 5 | 6 | 56 | 56 | 357 | 1190 | 3645 | 10.04 | 10.33 |
| 28 | 3 | 8 | 66 | 66 | 350 | 1190 | 3294 | 11.08 | 10.82 |
| 29 | 5 | 6 | 56 | 56 | 343 | 1190 | 3645 | 10.03 | 9.74 |
| 30 | 3 | 8 | 62 | 62 | 337 | 1190 | 3434 | 10.87 | 10.37 |
| 31 | 3 | 8 | 60 | 60 | 330 | 1190 | 3504 | 10.77 | 10.64 |
| 32 | 1 | 10 | 62 | 62 | 323 | 1190 | 3434 | 11.42 | 11.21 |
| 33 | 10 | 1 | 62 | 62 | 323 | 1190 | 3434 | 8.90 | 9.33 |
| 34 | 8 | 3 | 60 | 60 | 330 | 1190 | 3504 | 9.37 | 9.37 |
| 35 | 8 | 3 | 62 | 62 | 337 | 1190 | 3434 | 9.47 | 9.50 |
| 36 | 6 | 5 | 56 | 56 | 343 | 1190 | 3645 | 9.75 | 10.08 |
| 37 | 8 | 3 | 66 | 66 | 350 | 1190 | 3294 | 9.68 | 9.04 |
| 38 | 6 | 5 | 56 | 56 | 357 | 1190 | 3645 | 9.76 | 9.75 |
| 39 | 6 | 5 | 62 | 62 | 363 | 1190 | 3434 | 10.06 | 9.90 |
| 40 | 4 | 7 | 48 | 48 | 370 | 1190 | 3925 | 9.95 | 10.45 |
| 41 | 8 | 3 | 74 | 74 | 377 | 1561 | 2643 | 10.13 | 9.71 |
| 42 | 6 | 5 | 56 | 56 | 383 | 1561 | 3274 | 9.83 | 10.32 |
| 43 | 6 | 5 | 66 | 66 | 0 | 1561 | 1723 | 9.76 | 10.19 |
| 44 | 4 | 7 | 44 | 44 | 0 | 1561 | 2495 | 9.27 | 10.01 |
| 45 | 6 | 5 | 74 | 74 | 0 | 1260 | 1743 | 10.12 | 10.52 |
| 46 | 4 | 7 | 48 | 48 | 0 | 1260 | 2655 | 9.43 | 9.95 |
| 47 | 4 | 7 | 62 | 62 | 0 | 1260 | 2164 | 10.10 | 10.67 |
| **48** | **2** | **9** | **32** | **32** | **0** | **1260** | **3217** | **9.22** | **16.56** |
| 49 | 8 | 3 | 90 | 90 | 0 | 1030 | 1578 | 10.33 | 9.86 |
| 50 | 6 | 5 | 56 | 56 | 0 | 1030 | 2771 | 9.25 | 9.89 |
| 51 | 6 | 5 | 74 | 74 | 0 | 1030 | 2139 | 10.12 | 10.07 |
| 52 | 4 | 7 | 36 | 36 | 0 | 1030 | 3472 | 8.85 | 9.67 |
| 53 | 6 | 5 | 86 | 86 | 0 | 1100 | 1648 | 10.70 | 11.05 |
| **54** | **4** | **7** | **44** | **44** | **0** | **1100** | **3121** | **9.24** | **38.49** |
| **55** | **4** | **7** | **66** | **66** | **0** | **800** | **2650** | **10.27** | **15.76** |
| 56 | 2 | 9 | 20 | 20 | 0 | 800 | 4264 | 8.62 | 9.50 |
| 57 | 6 | 5 | 98 | 98 | 0 | 870 | 1623 | 11.28 | 10.41 |
| 58 | 4 | 7 | 48 | 48 | 0 | 870 | 3377 | 9.44 | 10.06 |

| 59 | 4 | 7 | 74 | 74 | 0 | 940 | 2395 | 10.70 | 10.57 |
| 60 | 2 | 9 | 20 | 20 | 0 | 640 | 4590 | 8.63 | 12.63 |
| 61 | 4 | 7 | 90 | 90 | 0 | 710 | 2229 | 11.47 | 12.13 |
| 62 | 2 | 9 | 32 | 32 | 0 | 480 | 4494 | 9.22 | 9.22 |
| 63 | 2 | 9 | 62 | 62 | 0 | 320 | 3767 | 10.67 | 10.45 |
| 64 | 0 | 11 | 0 | 0 | 0 | 0 | 6428 | 8.24 | 9.24 |

Note: defective optical paths are highlighted with red in the table.

However, the excess losses of Path $T_{i,j}$ ( $i = j$ = 5, 21, 22, 48, 54, 55, marked in red in Table 2) are significantly higher than the others. To investigate the higher excess loss of these 6 optical paths, we checked them using SEM and optical microscope. Some fabrication defects were found in the 6 optical paths, as shown in Fig. S18. The 6 defective optical paths are highlighted with red, blue, orange, green, yellow and purple, respectively, and locations of the defects are marked by stars with the same color in Fig. S18a. The transmission spectra of the 6 defective paths and the microscope images with a SEM close-up view of the corresponding defects are shown in Fig. S18b – g.

The main contributors to the excess loss of the switch array include switch cell loss, waveguide transmission loss, waveguide bending loss, and waveguide crossing loss. In Section 11, the excess losses of Euler S-bends (26 mdB) and varied-width waveguide crossings (28 mdB) are measured without excluding the propagation loss. As a result, they can be further reduced by improving the fabrication capabilities for lower propagation loss as well. The arc length of Euler S-bend is 62.8 μm, suggesting that 18 mdB of the 26 mdB loss is due to propagation loss (2.8 dB/cm). The measured excess loss of crossing is 28 mdB but only 4 mdB in theory, which is caused by the propagation loss as well. In our previous work, we have achieved varied-width waveguide crossings with measured excess loss of 6 mdB[16]. Therefore, by using waveguides with lower propagation loss with improved fabrication capabilities, the excess losses of Euler S-bend and waveguide crossings can be reduced to < 10 mdB. If we retain the switch design in this paper ($T_{11}$ = 0.36 dB, $T_{22}$ = 0.65 dB) and improve the fabrication to achieve 0.1 dB/cm propagation loss for 1.2 μm-wide waveguide as well as 10 mdB excess loss for Euler S-bends and waveguide crossings, the excess loss for the 64 × 64 Benes switch array can be reduced to 4.1 – 8.2 dB (6.3 dB in average). When it is desired to scale the switch array up to 128×128, the excess loss would be approximately 4.9–11.4 dB (9.1 dB in average). It can be seen that the key to break the bottleneck for scaling is the fabrication improvement.

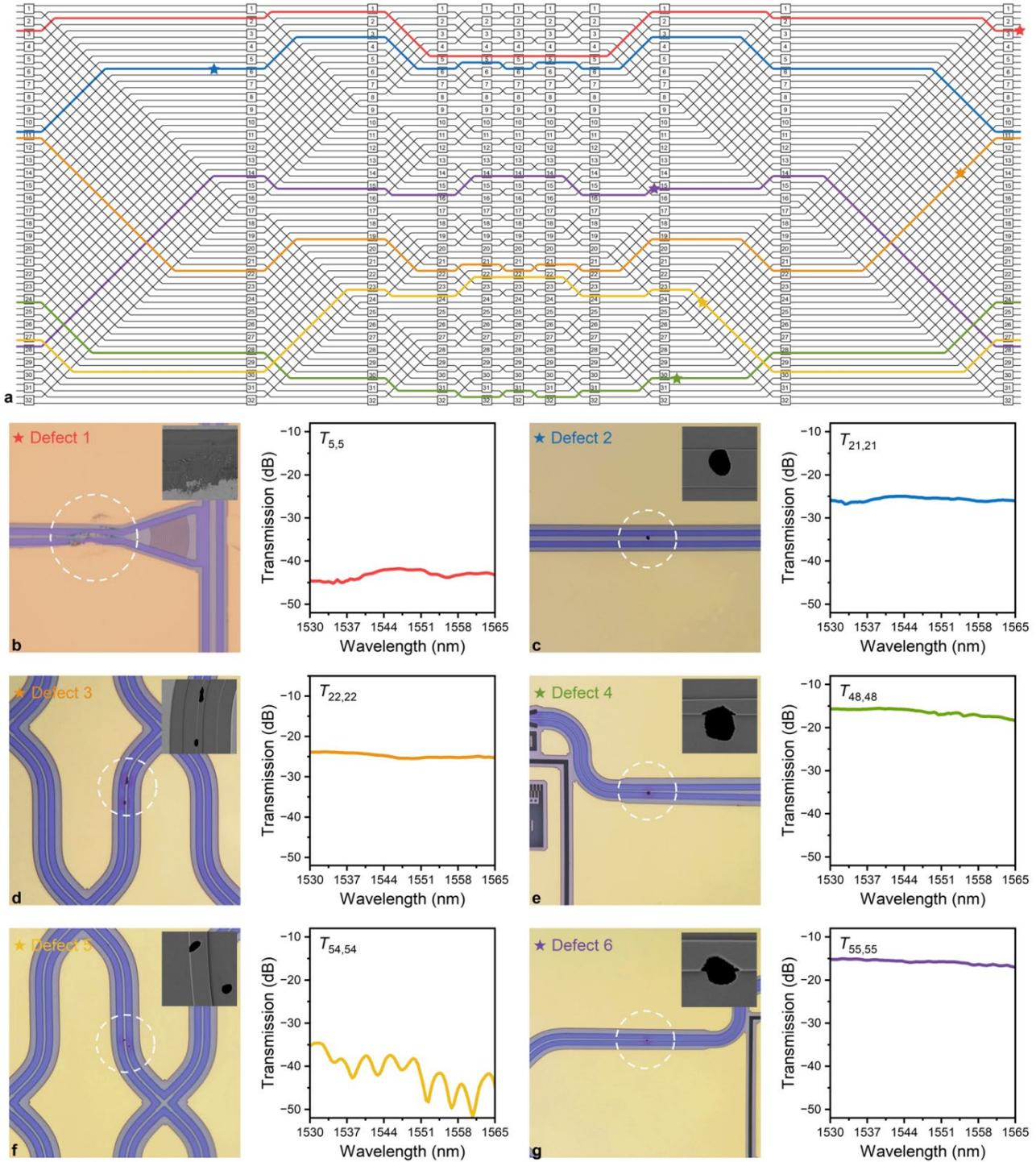

**Fig. S18 | Analysis of the defective paths in all-OFF state. a,** Schematic diagram of the switch array, the defective paths $T_{i,j}$ ($i = j = 5, 21, 22, 48, 54, 55$) are highlighted with red, blue, orange, green, yellow and purple, respectively. Locations of the defects are marked by stars. **b − g,** Optical microscope images and SEM close-up views of the defects and the transmission spectra of the corresponding defective paths, respectively.

Switch cells in the center stage of the 64 × 64 Benes switch array were actuated to reconfigure the optical path $T_{i,j}$ ($i = j = 1 − 32$) as shown in Table 3. In a Benes topology, when the switch cells in the center stage are turned on, optical paths $T_{i,j}$ ($i = j = 1 − 32$) will be reconfigured to $T_{i,j}$ ($i = 1 − 32, j = i + 32$) and the reconfigured path $T_{i,j}$ coincides with the left half of path $T_{i,i}$ (all-OFF state) and the right half of path $T_{j,j}$ (all-OFF state). We turned on the switch cells in the center stage sequentially and the ON/OFF transmission spectra of the corresponding output ports are measured. The excess losses of the reconfigured optical paths at 1550 nm are listed in Table 3.

Table 3. Measured results of the fabricated 64 × 64 Benes switch array with a single switch actuated. (single-ON)

| Initial path | Reconfigured path | Actuated switch index | Output port index | Excess Loss (dB) | Actuation voltage (V) |
|---|---|---|---|---|---|
| $T_{1,1}$ | $T_{1,33}$ | 6-1 | 33 | -10.4905 | 20 V |
| $T_{17,17}$ | $T_{17,49}$ | 6-2 | 49 | -11.9818 | 20 V |
| $T_{9,9}$ | $T_{9,41}$ | 6-3 | 41 | -9.26884 | 20 V |
| $T_{25,25}$ | $T_{25,57}$ | 6-4 | 57 | -9.04217 | 20 V |
| $T_{5,5}$ | $T_{5,37}$ | 6-5 | 37 | -12.5551 | 20 V |
| $T_{21,21}$ | $T_{21,53}$ | 6-6 | 53 | -42.7658 | 20 V |
| $T_{13,13}$ | $T_{13,45}$ | 6-7 | 45 | -12.9598 | 20 V |
| $T_{29,29}$ | $T_{29,61}$ | 6-8 | 61 | -12.4817 | 20 V |
| $T_{3,3}$ | $T_{3,35}$ | 6-9 | 35 | -10.0526 | 20 V |
| $T_{19,19}$ | $T_{19,51}$ | 6-10 | 51 | -10.3446 | 20 V |
| $T_{11,11}$ | $T_{11,43}$ | 6-11 | 43 | -12.8573 | 20 V |
| $T_{27,27}$ | $T_{27,59}$ | 6-12 | 59 | -10.5668 | 20 V |
| $T_{7,7}$ | $T_{7,39}$ | 6-13 | 39 | -10.6523 | 20 V |
| $T_{23,23}$ | $T_{23,55}$ | 6-14 | 55 | -15.8619 | 20 V |
| $T_{15,15}$ | $T_{15,47}$ | 6-15 | 47 | -11.8102 | 20 V |
| $T_{31,31}$ | $T_{31,63}$ | 6-16 | 63 | -9.52457 | 20 V |
| $T_{2,2}$ | $T_{2,34}$ | 6-17 | 34 | -9.74949 | 20 V |
| $T_{18,18}$ | $T_{18,50}$ | 6-18 | 50 | -9.95807 | 20 V |
| $T_{10,10}$ | $T_{10,42}$ | 6-19 | 42 | -11.1695 | 20 V |
| $T_{26,26}$ | $T_{26,58}$ | 6-20 | 58 | -9.08256 | 20 V |
| $T_{6,6}$ | $T_{6,38}$ | 6-21 | 38 | -11.9261 | 20 V |
| $T_{22,22}$ | $T_{22,54}$ | 6-22 | 54 | -38.6567 | 20 V |
| $T_{14,14}$ | $T_{14,46}$ | 6-23 | 46 | -11.8879 | 20 V |
| $T_{30,30}$ | $T_{30,62}$ | 6-24 | 62 | -9.84546 | 20 V |
| $T_{4,4}$ | $T_{4,36}$ | 6-25 | 36 | -12.7724 | 20 V |
| $T_{20,20}$ | $T_{20,52}$ | 6-26 | 52 | -11.0261 | 20 V |
| $T_{12,12}$ | $T_{12,44}$ | 6-27 | 44 | -12.7096 | 20 V |
| $T_{28,28}$ | $T_{28,60}$ | 6-28 | 60 | -10.5591 | 20 V |
| $T_{8,8}$ | $T_{8,40}$ | 6-29 | 40 | -13.2957 | 20 V |
| $T_{24,24}$ | $T_{24,56}$ | 6-30 | 56 | -9.39372 | 20 V |
| $T_{16,16}$ | $T_{16,48}$ | 6-31 | 48 | -22.416 | 20 V |
| $T_{32,32}$ | $T_{32,64}$ | 6-32 | 64 | -11.6222 | 20 V |

Note: defective optical paths are highlighted with red in the table.

Higher excess losses of reconfigured paths $T_{21,53}$, $T_{23,55}$, $T_{22,54}$ and $T_{16,48}$ can be observed.

The 4 reconfigured paths are highlighted with red, blue, orange and green in Fig.S19a. Locations of the 6 defects described in Fig. S18 are marked with stars in Fig.S19a as well. Evidently, the higher excess losses of the 4 reconfigured paths are caused by the same defects shown in Fig. S18. Defective reconfigured paths $T_{21,53}$, $T_{23,55}$, $T_{22,54}$ and $T_{16,48}$ correspond to the defects 2, 6, 5, 4 respectively. The transmission spectra of the 4 defective reconfigured paths are shown in Fig.S19b – e.

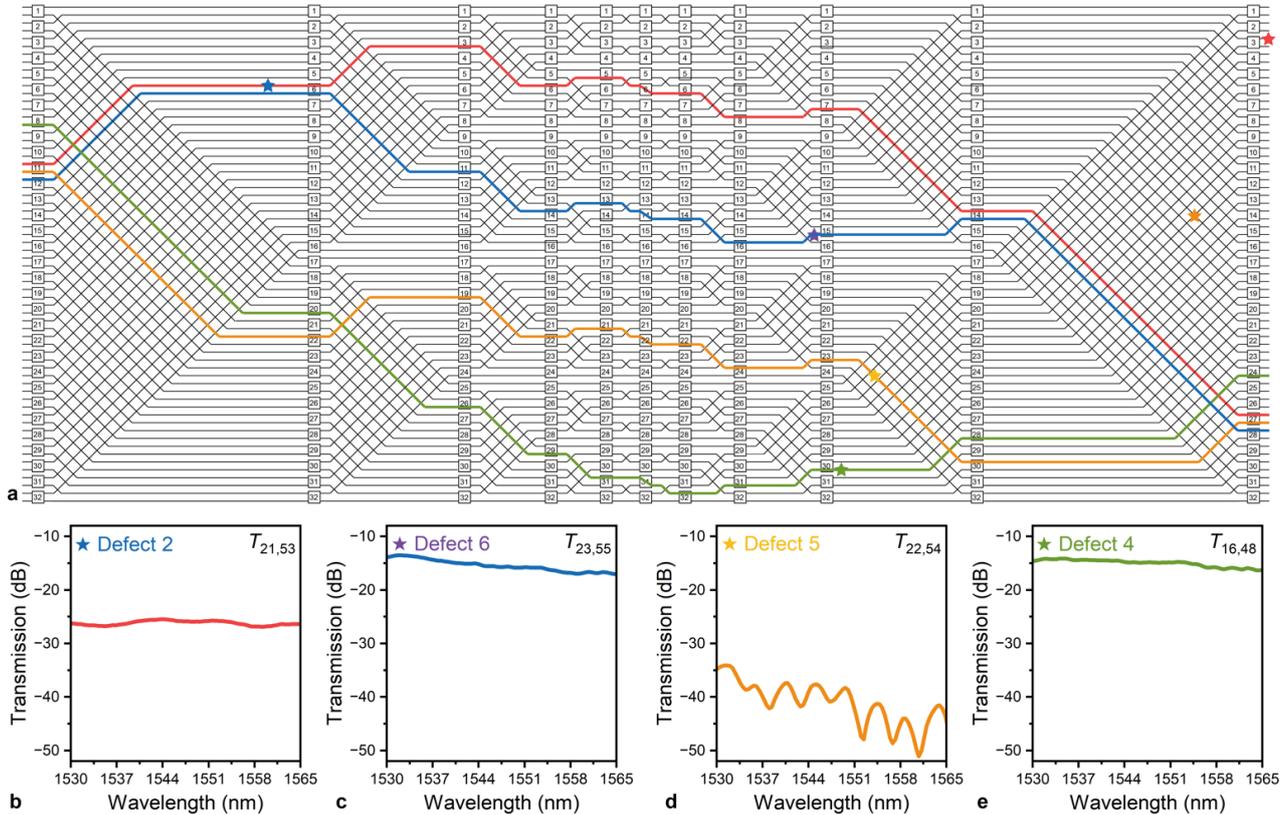

**Fig.S19 | Analysis of the defective paths in single-ON state. a,** Schematic diagram of the switch array, the defective reconfigured paths $T_{21,53}$, $T_{23,55}$, $T_{22,54}$ and $T_{16,48}$ are highlighted with red, blue, orange and green, respectively. **b – e,** The transmission spectra of the corresponding defective paths, respectively.

## Section 13: Measurement Setup

The measurement setups of the 2 × 2 elementary switch and the 64 × 64 Benes switch are shown in Figs. S20 and S21, respectively. During all the testing process, an amplified spontaneous emission laser (1520–1610 nm) was used as the light source and the drive voltage signals were applied to the devices through direct-current probes. All the optical transmission spectra were measured with an optical spectrum analyzer (OSA, Yokogawa, AQ6370D).

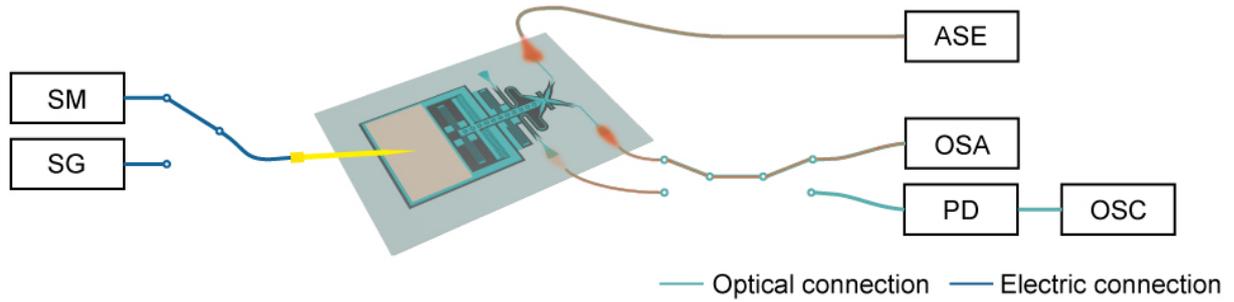

Fig. S20 | Schematic illustration of the measurement setup for an elementary switch.

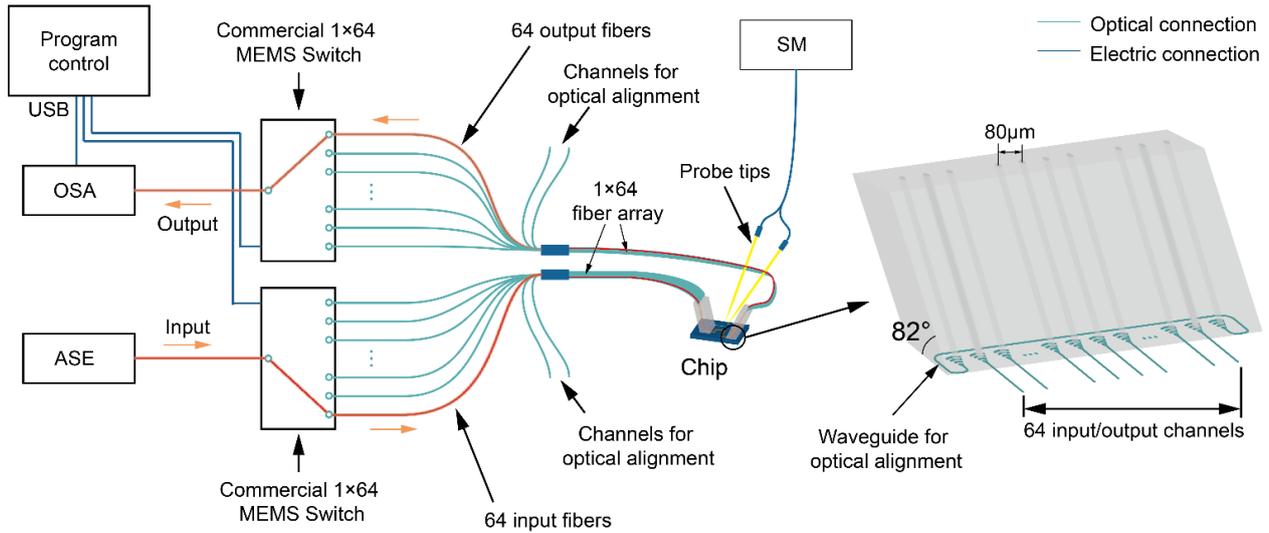

Fig. S21 | Schematic illustration of the measurement setup for the 64 × 64 Benes switch.

A pair of fibers were used to characterize the performance of the elementary switch. The fiber-to-chip coupling loss was measured to be 5 dB. In the ON/OFF optical transmission spectra test, the drive voltage was generated by a sourcemeter (Keithley 2400). The square-wave voltage for the temporal response measurement and the durability measurement was generated by a signal generator (SIGLENT SDG1032X). In the temporal response measurement, light output from the switch was measured with a photodiode (Newport 2053) and an oscilloscope (Tektronix, TBS1000C)

A pair of 66-channel fiber array (spaced 80 µm, angled 8°) was used to characterize the performance of the 64×64 Benes switch. To measure the 4096 transmission spectra of the Benes switch in its initial (all-OFF) state, a pair of commercial 1×64 optical switches were used to perform automatic measurements. The commercial 1×64 optical switches and the OSA were controlled by a MATLAB program to achieve automatic light path switching and optical transmission spectrum measurement.